
\input harvmac
\noblackbox

\def \FM {$F$-model\  }
\def \KM {$K$-model\  }

\def \eq#1 {\eqno {(#1)}}

\def\bl{\left}
\def\br{\right}

\def \ra {\rightarrow}
\def\np {  Nucl. Phys. }
\def \pl { Phys. Lett. }
\def \mpl { Mod. Phys. Lett. }
\def \prl { Phys. Rev. Lett. }
\def \pr  { Phys. Rev. }

\def \ijmp { Int. J. Mod. Phys. }

\def\k{\kappa}
\def\r{\rho}
\def\a{\alpha}
\def\b{\beta}
\def\B{\Beta}
\def\g{\gamma}

\def\d{\delta}

\def\e{\epsilon}

\def\p{\phi}

\def\th{\theta}

\def\m{\mu}
\def\n{\nu}

\def\l{\lambda}

\def\s{\sigma}

\def \sm {$\s$-model\ }

\def \bd {\bar \del}

\def \A { {\bar A} }

\def \ha {{1\over 2}}

\def \ov {\over}

\def\const{{\rm const}}
\def \p {\phi}
\def \vp {\varphi}

\def\bl{\bigl}
\def\br{\bigr}

\def \bd {\bar \del}

\def \tt {{\tilde \t}}

\def \O {\Omega }

\def \ra {\rightarrow}

\def \na {\nabla }

\def \a {\alpha}
\def \b {\beta}

\def \Tr {{\ \rm Tr \ }}

\def \ln {{\rm \ ln \  }}
\def \det {{\ \rm det \ }}
\def \ch {{\rm cosh \ }}
\def \th {{\rm tanh  }}
\def \l {\lambda}
\def \p {\phi}

\def \m {\mu }
\def \n {\nu}
\def \ep {\epsilon}
\def\g {\gamma}
\def \r {\rho}
\def \k {\kappa }
\def \d {\delta}
\def \o {\omega}
\def \s {\sigma}
\def \t {\theta}

\def \fourth {{\textstyle{1\over 4}}}

\def \e#1 {{{\rm e}^{#1}}}
\def \const {{\rm const }}

\def \eq#1 {\eqno {(#1)}}
\def \sm {$\s$-model\ }\def \B  {{ \tilde B }}

\def \bd  {{ \bar \del }}

\def \bd  { \bar \del }

\def \A { \bar A}


\def \o {\omega}

\def \p {\phi}
\def \ep {\epsilon}
\def \s {\sigma}

\def \r {\rho}
\def \d {\delta}
\def \l {\lambda}
\def \m {\mu}

\def \g {\gamma}
\def \n {\nu}

\def \fourth {{1\over 4}}

\def \e#1 {{{\rm e}^{#1}}}
\def \const {{\rm const }}\def \vp {\varphi}
\def \B {{\bar B}}

\def \m {\mu}  

\def \ep {\epsilon}

\def \ra {\rightarrow}

\def \const {{\rm const} }

\def \eq#1 {\eqno{(#1)}}
\def \e {\rm e}
\def \ra {\rightarrow }
\def \e#1 {{\rm e}^{#1}}

\def \ch {\ {\rm cosh} \ }
\def \th {\ {\rm tanh} \ }
\def \ln { {\rm ln } }

\def \l {\lambda}
\def \p {\phi}
\def \vp {\varphi}
\def  \g {\gamma}
\def \o {\omega}
\def \r {\rho}

\def\({\left (}
\def\){\right )}
\def\[{\left [}
\def\]{\right ]}

\def\np {  Nucl. Phys. }
\def \pl { Phys. Lett. }
\def \mpl { Mod. Phys. Lett. }
\def \prl { Phys. Rev. Lett. }
\def \pr  { Phys. Rev. }
\def \cqg { Class. Quantum Grav.}

\baselineskip8pt
\Title{\vbox
{\baselineskip 6pt{\hbox{  }}{\hbox
{Imperial/TP/93-94/38 }}{\hbox {NI94004}}{\hbox{hep-th/9406067}} } }
{\vbox{\centerline {  On exact  solutions and singularities in string theory}
 }}
\vskip 37 true pt
\centerline  { {Gary T. Horowitz\footnote {$^*$} {e-mail address:
gary@cosmic.physics.ucsb.edu} }\footnote {$^{**}$}{On leave from Physics
Department, University of California, Santa Barbara, CA 93106, USA}}
 \smallskip \smallskip
\centerline {\it  Isaac Newton Institute }
\smallskip

\centerline{\it Cambridge CB3 0EH, U.K.}
\medskip
\centerline {and}
\medskip
\centerline{   A.A. Tseytlin\footnote{$^{\star}$}{\baselineskip8pt
e-mail address: tseytlin@ic.ac.uk}\footnote{$^{\dagger}$}{\baselineskip5pt
On leave  from Lebedev  Physics
Institute, Moscow, Russia.} }

\smallskip\smallskip
\centerline {\it  Theoretical Physics Group, Blackett Laboratory}

\centerline {\it  Imperial College,  London SW7 2BZ, U.K. }
\medskip
\centerline {\bf Abstract}
\smallskip
\baselineskip6pt
\noindent
We construct two new classes of exact solutions to string theory which
are not of the standard plane wave or gauged WZW type. Many of these solutions
have curvature singularities. The first class includes the fundamental
string solution, for which the string coupling vanishes near the
singularity. This suggests that the singularity may not be removed by
quantum corrections. The second class consists of  hybrids of plane
wave and gauged WZW solutions. We discuss a four dimensional example
in detail.

\Date {June  1994}

\noblackbox
\baselineskip 16pt plus 2pt minus 2pt

\vfill\eject

\def \lr { \lref}

\gdef \jnl#1, #2, #3, 1#4#5#6{ {\sl #1~}{\bf #2} (1#4#5#6) #3}

\lr \mans { P. Mansfield and J. Miramontes, \jnl \pl, B199, 224, 1988;
A. Tseytlin, \jnl \pl, B208, 228, 1988; \jnl \pl, B223, 165, 1989.}

\lr \kalmor{R. Kallosh and A. Morozov,  \jnl \ijmp,  A3, 1943, 1988.}

\lr \ghrw{J. Gauntlett, J. Harvey, M. Robinson, and D. Waldram,
\jnl \np, B411, 461, 1994.}
\lr \garf{D. Garfinkle, \jnl \pr, D46, 4286, 1992.}

\lr \onofri { V. Fateev, E. Onofri and Al. Zamolodchikov, \jnl \np, B406,
 521, 1993.}

\lref \tspl {A. Tseytlin, \jnl \pl, B317, 559, 1993.}
\lref \tssfet { K. Sfetsos and A.  Tseytlin, \jnl  \pr, D49, 2933, 1994.}
\lref \klts {C. Klim\v c\'\i k  and A. Tseytlin, ``Exact four dimensional
string solutions and Toda-like sigma models from null-gauged
WZNW models",  preprint
 Imperial/TP/93-94/17, hep-th/9402120.}

\lr \sfexac {K. Sfetsos,  \jnl \np, B389, 424,  1993.}

\lr \tsmac{A. Tseytlin, \jnl \pl,  B251, 530, 1990.}

\lr \cakh{C. Callan and R. Khuri, \jnl \pl, B261, 363, 1991;
R. Khuri, \jnl \np, B403, 335, 1993.}
\lr \dgt{M. Duff, G. Gibbons and P. Townsend, ``Macroscopic superstrings
as interpolating solitons", DAMTP/R-93/5, hep-th/9405124.}

\lref \ger {A. Gerasimov, A. Morozov, M. Olshanetsky, A. Marshakov and S.
Shatashvili, \jnl \ijmp,
A5, 2495,  1990. }

\lr \hutow{C. Hull and P. Townsend, \jnl \np, B274, 349, 1986.}
\lr \mukh {S. Mukhi,    \jnl \pl,  B162, 345, 1985;
S. De Alwis, \jnl \pl, B164, 67, 1985. }

\lr \napwietc { C. Nappi and E. Witten, \jnl \prl, 71, 3751, 1993;
E. Kiritsis and C. Kounnas, \jnl \pl, B320, 264, 1994; D.  Olive,
 E. Rabinovici and A. Schwimmer,  \jnl \pl, B321, 361, 1994;
 K. Sfetsos,  \jnl \pl, B324, 335, 1994;
  ``Gauged WZW models and Non-abelian Duality'',
THU-94/01, hep-th/9402031;
I. Antoniadis and N. Obers, ``Plane Gravitational Waves in String
Theory'', CPTH-A299.0494, hep-th/9403191;
K. Sfetsos and A. Tseytlin,  ``Four Dimensional Plane Wave String Solutions
with Coset CFT Description", preprint THU-94/08, hep-th/9404063. }

\lr \scherk { J. Scherk, \jnl \np, B31,  222, 1971;
   J. Scherk and J. Schwarz, \jnl \np, B81, 118,
1974;  T. Yoneya, \jnl {\it Progr. Theor. Phys.}, 51, 1907, 1974.}
 \lr \lov  {C. Lovelace,  \jnl \np,  B273, 413,  1986.}
\lr \call{C. Callan, D. Friedan, E. Martinec and  M. Perry, \jnl \np, B262,
593, 1985.}
\lr \frts {E.  Fradkin  and A. Tseytlin, \jnl \pl, B158, 316, 1985;
\jnl \np, B261, 1, 1985.}
\lr \tsred  {A. Tseytlin, \jnl  \pl, B176, 92, 1986; \jnl  \np, B276, 391,
 1986.}

\lr \gps {S.  Giddings, J. Polchinski and A. Strominger, \jnl  \pr,  D48,
 5784, 1993. }

\lr \tsppl  {A. Tseytlin, \jnl   \pl,  B208, 221, 1988.}
\lr\rabi  {S. Elitzur, A. Forge and E. Rabinovici, \jnl \np, B359, 581, 1991;
 G. Mandal, A. Sengupta and S. Wadia, \jnl \mpl,  A6, 1685, 1991. }
 \lr \witt{ E. Witten, \jnl \pr, D44, 314, 1991. }
 \lr \dvv { R. Dijkgraaf, H. Verlinde and E. Verlinde, \jnl \np, B371,
269, 1992.}
\lr \hoho { J. Horne and G.  Horowitz, \jnl \np, B368, 444, 1992. }
\lr \horwel{G. Horowitz and D. Welch, \jnl \prl, 71, 328, 1993;
N. Kaloper,  \jnl \pr,  D48, 2598, 1993. }
\lr \host{ G. Horowitz and A. Steif,  \jnl \prl, 64, 260, 1990; \jnl \pr,
D42, 1950, 1990;  G. Horowitz, in: {\it
 Strings '90}, (eds. R Arnowitt et. al.)
 World Scientific (1991).}
\lr \busch {T.  Buscher, \jnl \pl, B194, 59, 1987; \jnl \pl,
 B201, 466, 1988.}
\lr \kallosh {E. Bergshoeff, I. Entrop, and R. Kallosh, ``Exact Duality in
String Effective Action", SU-ITP-93-37; hep-th/9401025.}

\lr \tsmpl {A. Tseytlin, \jnl  \mpl, A6, 1721, 1991.}
\lr \vene { }
\lr \kltspl { C. Klim\v c\'\i k and A. Tseytlin, \jnl \pl, B323, 305, 1994.}
\lr \shwts { A. Schwarz and A. Tseytlin, \jnl \np, B399, 691, 1993.}
\lr \callnts { C. Callan and Z. Gan, \jnl  \np, B272, 647, 1986;  A. Tseytlin,
\jnl \pl,
B178, 34, 1986.}

\lr \guv { R. G\"uven, \jnl \pl, B191, 275, 1987;
 D. Amati and C. Klim\v c\'\i k,
\jnl \pl, B219, 443, 1989;
 R. Rudd, \jnl \np, B352, 489, 1991.}
\lr \desa{ H. de Vega and N. Sanchez, \jnl
\pr, D45, 2783, 1992; \jnl \cqg, 10, 2007, 1993.}
\lr \desas{ H. de Vega and N. Sanchez, \jnl
\pl, B244,  215, 1990.}
\lref \tsnul { A. Tseytlin, \jnl \np, B390, 153, 1993.}

\lref \dunu { G. Horowitz and A. Steif, \pl B250 (1990) 49;
 E. Smith and J. Polchinski, \pl B263 (1991) 59. }

\lr \gauged {I. Bars and K. Sfetsos, \jnl  \mpl, A7, 1091, 1992;
 P. Horava, \jnl \pl,
B278, 101, 1992; P. Ginsparg and F. Quevedo, \jnl \np, B385, 527, 1992. }
\lr \bsfet {I. Bars and K. Sfetsos, \jnl \pr, D46, 4510, 1992; \jnl \pr,
 D48, 844, 1993. }
\lr \tsnp{ A. Tseytlin, \jnl \np, B399, 601, 1993;  \jnl \np, B411, 509, 1994.}
\lr \gibb{A. Dabholkar, G. Gibbons, J. Harvey, and F. Ruiz, \jnl \np, B340,
33, 1990.}
\lr \hhs{J. Horne, G. Horowitz, and A. Steif, \jnl \prl, 68, 568, 1992;
G. Horowitz, in:  {\sl String Theory and
Quantum Gravity '92}, (eds. J. Harvey, R. Iengo, K. Narain, S. Randjbar-Daemi,
and H. Verlinde) World Scientific (1993).}

\lr \polypolchnats {  }
\lr \jack {I. Jack, D.  Jones and J. Panvel, \jnl \np, B393, 95, 1993.}
\lr \mettstwo {R.  Metsaev and A. Tseytlin, \jnl \pl, B185, 52, 1987.}
\lr \banks {T. Banks, M. Dine, H. Dijkstra and W. Fischler, \jnl \pl, B212,
45, 1988.}
\lr \horstr{G. Horowitz and A. Strominger, \jnl \np, B360, 197, 1991.}

\lr \givkir {A.  Giveon and E. Kiritsis, \jnl \np, B411, 487, 1994.  }

\lr \jac{I. Jack and D. Jones, \jnl \pl, B200, 453, 1988.}
\lr \metts{R. Metsaev and A. Tseytlin, \jnl \np,  B293, 385, 1987.  }

\newsec{Introduction}

One of the main obstacles toward a better understanding of string theory
is the scarcity of exact classical solutions. At the present time, only
two classes of solutions are known for the bosonic string. The first
are
plane-wave-type  backgrounds \refs{\guv,\host,\desa,\tsnul}
which have a covariantly
constant null vector, and the second are those
corresponding to  gauged WZW  models (see e.g. \refs{\witt,\gauged}).
(For the superstring, there is a  third class of solutions corresponding
to $(2,2)$ supersymmetric models.)

There is a well known ambiguity in the form of the classical string
equations of motion. These equations are  usually expressed as a power series
in $\a'$. The leading term is unambiguous,
 but the form of the higher
order terms can be altered by field redefinitions or
equivalently, by choosing different `renormalization schemes'. For the simplest
plane wave  solutions this ambiguity is irrelevant since all the higher order
terms vanish identically. For the gauged WZW solutions, in
the familiar conformal field theory (CFT) scheme there  are
$\a'$-corrections  to all orders \refs{\dvv,\bsfet,\tsnp},
 but there  is evidence
\refs{\tspl,\tssfet,\givkir}  that there also
 exists a  scheme
where the leading-order  solution is exact.
 Given this ambiguity,
to study string propagation and scattering one needs to know more than
the fact that a particular background is exact in a certain scheme. One
also needs to identify explicitly the corresponding
CFT. This is known for the gauged
WZW models, but not for all the plane wave
solutions. Nevertheless, some properties of a solution can
be determined from the information about its exact form in some scheme.

Is it possible to  go beyond these two classes of solutions? Consider
a family of backgrounds described by a dilaton $\p (x)$ and a
metric and antisymmetric tensor characterized by a single function $F(x)$:
\eqn\ffff{ ds^2 = F(x) du dv + dx_i dx^i\ ,   \qquad B_{uv} = \ha F(x) \ . }
Note that the two functions $F$ and $\p$ depend only
on the transverse coordinates
$x^i$. For backgrounds of this form, the leading order equations of motion
reduce to  (see Appendix A)
\eqn\fmod{ \del^2 F\inv = 2b^i \del_i F\inv\ , \ \ \
  \p = \p_0 + b_i x^i  + \ha \ln F (x) \ , \ }
where $b_i$ is a  constant vector.
Some of the solutions to \fmod\ have recently been shown   to
correspond to gauged WZW models where the subgroup being gauged
 is nilpotent \klts. It was
argued that they should not receive higher order corrections in
the CFT scheme.
It is currently
unknown whether all solutions to \fmod\ can be obtained from a gauged WZW model
but we believe this to be unlikely. Nevertheless, we will show that there
is a scheme in which all of these solutions are exact
and receive no $\a'$ corrections. Since the equation for $F\inv$ is linear,
linear combinations of these solutions yield new exact solutions.

One of the most interesting solutions in this class  is the
fundamental string (FS) \gibb\
 which
has $b_i=0$ and
\eqn\fss{ F\inv ={ 1 + {M \ov r^{D-4}} }   \ , \ \ D>4 \ ;}
$$
 F\inv ={ 1 - {M \  \ln \ r } } \ , \ \ \  D=4\ , $$
where $r^2 = x_i x^i$ and $D$ is the number of spacetime dimensions.
This solution
describes the field outside of a straight
fundamental string located at
$r=0$.
The metric \ffff\  becomes degenerate at $r=0$ and the curvature diverges.
One would therefore expect
the higher order terms in the string equations to become important and
significantly modify the solution. However it was recently suggested
\kallosh\ that this might not be the case for the superstring. Some
evidence based on supersymmetry was presented that the FS does not
receive higher order $\a' $ corrections.\foot{ It was shown in
 \kallosh\  that corrections to the equations of motion coming
 from specific (anomaly-related) terms in the effective action vanish
 on the FS background.
However,  their argument is incomplete since
the heterotic string effective   action also contains other terms
(necessary for reproducing the string $S$-matrix)  which were not
considered in \kallosh.
It may be that the contributions of these other terms (taken in a specific
`supersymmetric'  scheme)  also  vanish on  the FS background but
this question deserves further  investigation.}
One consequence of our results is that  the FS is indeed
an exact solution (in a particular scheme) even for the bosonic  string theory,
and thus of course is a heterotic or superstring solution as well.

It is known that the FS in $D>4$ is the extremal limit of a two
parameter  charged black string solution to the
leading order equations  with  regular
event horizon \refs{\horstr,\hhs}. The nonextremal black string receives
higher order
$\a'$ corrections. But in
$D=3$ a two-parameter charged black string solution was constructed
from a gauged WZW model \hoho.  It is likely that  there  exists a scheme
in which  this leading order
solution is exact (this  was shown   to order $\a'$ in \tssfet). Its extremal
limit
turns out the be the general
  solution to \fmod\ in three dimensions. This provides a perturbative
check on the general argument for the conformal invariance of these
backgrounds.
In the original construction, the extremal $D=3$  black string was obtained
by taking a certain limit of a background representing a  gauged WZW model.
We will show that this extremal solution
can be obtained directly as a particular
 gauged  $SL(2,R)\times R/R$ WZW model.

The fact that the FS \fss\ is an exact classical solution has implications for
singularities in string theory.
We do not yet have a completely satisfactory definition of  a singularity
in classical string theory. Geodesic incompleteness, which is so useful
in general relativity, is clearly unsatisfactory as seen e.g. by
 the example  of orbifolds.
Even diverging curvature   is not a sufficient condition since some solutions
with curvature singularities are known to be equivalent to nonsingular
backgrounds. A definition based on the motion of classical strings
is no better
than geodesics, since null geodesics are included in the motion of strings.
It appears that one must define a singularity in terms of the motion of
quantum strings. This is natural since the equations for the background
fields come from requiring that quantum strings are described by a
CFT. Thus a string singularity should be a
CFT which is  ill-behaved in some sense. For the simplest plane
wave solutions, one can study the propagation of quantum strings
explicitly and show that this is not well behaved when the wave
becomes singular \host\ (see also \desa). So at least some
 classical singularities exist in
string theory.

We do not yet know if the CFT associated with the FS is ill-behaved due to
the curvature singularity.  If it is,  this may
have a striking consequence.
	 One usually expects
that quantum effects will be large near regions of large curvature.
But the string coupling is a dynamical field $e^\p$, and  for the FS solution
$e^\p \rightarrow 0$ near the singularity. This is quite different from
other familiar examples of classical solutions such as the two dimensional
black hole \witt\ for which $e^\p $ diverges at the singularity in the leading
order metric. (If one just considers solutions to the leading order equations,
one can obtain a large class of singular backgrounds by starting with any
 regular solution with a symmetry having a fixed point,
 and applying a spacetime duality transformation
\busch. In all these examples, $e^\p $ diverges at the singularity.)
The fact that $e^\p \rightarrow 0$ for the FS suggests that quantum
loop corrections will be suppressed and the solution will become more
classical near the singularity. If this could  be established, it would
show that  at least  some  singularities remain even in quantum string theory.

The observation that all the leading order
  solutions of the form \ffff\ are  exact
 also has implications for
spacetime duality. These solutions can be obtained by applying a leading
order duality transformation (with respect to translations in $u$) to
the plane wave
metrics
\eqn\pwmet{ ds^2 = dudv +K(x) du^2 +dx^idx_i\ , }
with $K=F\inv$, \ $B_{\mu\nu} = 0$ and $\p = \p_0 + b_i x^i$. These solutions
are known to be exact.  One can ask whether there always exists a scheme
in which the leading order duality is not modified
by $\a'$-corrections. It turns out that  this is not the case:
the $\a'$-modification  \tsmpl\ of the leading-order duality
is necessary in all the schemes if $D\geq 3$. (A special scheme where duality
is not modified exists in $D=2$; this is not surprising since
as we shall see  the effective action is trivial in this case.)
However, it is possible that  the
following slightly weaker statement is true:
 Given an exact solution to string theory
with a continuous symmetry,
the solution obtained by a leading order duality transformation
is also exact in some scheme.

This conjecture does not require that duality itself be exact
since we allow the original solution and its dual to be exact in
different schemes. Some earlier evidence for this came from the fact
that the
 three dimensional
black hole constructed from the $SL(2,R)$ WZW model \horwel\ (which is exact)
is dual to the three dimensional
 black string \hoho, and it was shown \tssfet\ that the first $\a'$
correction to the black string metric can be removed by a field redefinition.
The fact that the leading order duals to the
plane wave solutions \pwmet\ also turn out to be exact is further
support for  this conjecture.

The method we will use to establish the conformal invariance of
\ffff\ also applies to a larger class of backgrounds where the
transverse space is curved. A similar argument can be  used to show
the conformal invariance of  the
plane wave
metrics \pwmet\ with  non-flat transverse part (see also \tsnul).
To obtain explicit
solutions, one needs the `transverse' theory to be, e.g.,  a gauged WZW
model.  In this way one obtains a `hybrid' of the plane wave and gauged WZW
solutions.

To illustrate this construction, we will discuss two examples.
 The smallest dimension for which the construction is nontrivial is $four$.
In this case we are able  to resolve a difficulty with the
FS in four dimensions.
The FS \fss\ in $D=4$ differs from its higher dimensional
analogs in that it has  an extra singularity at a
non-zero value of $r$. We will find that the dual of the
new solution we construct can be viewed as a fundamental string
in $D=4$ which is asymptotically flat and has no additional singularities.
We will also discuss a simple five dimensional example (which may also be
interpreted
as a $D=4$ heterotic string solution).

It is clear that when considering  exact solutions, the scheme dependence of
the equations of motion, or effective action (EA) that reproduces them,
plays an important role.
We shall show that in $D=2$  one can  actually represent the  $(G,B,\p)$
 part of the EA only by the leading-order term. In other words, all of
the higher order $\a'$-terms can be redefined away. In retrospect
this is not surprising for two reasons. First, in $D=2$,
the metric-dilaton system has no dynamical degrees of freedom
and the only propagating mode is a massless tachyon. So  there are
 no real massive exchanges and hence
no  genuine  $\a'$-vertices in the EA. Second, in $D=2$
the most general leading-order classical solution is the `black hole'
\refs{\witt, \rabi}.
This corresponds to the  $SL(2,R)/U(1)$ coset and  it is known that
 there exists a scheme \tspl\  where the leading order solution is not modified
by
higher-order corrections.

In $D=3$ one cannot remove all the higher order terms since the $(G,B,\p)$
system has one degree of freedom. However, we will see that
it is possible to choose
a scheme in which the higher order terms depend only on derivatives of
the dilaton.

The plan of this paper is as follows. In Section 2 we introduce the
general models we wish to consider, and derive the conditions under
which they are conformally invariant to all orders in $\a'$.
 In the next
section, we use these results to explicitly construct two new solutions,
one in four spacetime dimensions, and the other in five. Section 4
is devoted to a discussion of the field redefinition ambiguity and the
structure of the effective action in low dimensions. In Section 5
we will show that solutions to \fmod\  in $D=3$ correspond to
 a gauged WZW model.
Section 6 discusses the relation between solutions  (in particular, the ones
corresponding to gauged WZW models)
in different schemes.
Some concluding remarks are made in Section 7.  The appendices contain
some technical details and a further generalization of our models.

\newsec{Path integral argument for  conformal invariance}

\subsec{Basic  models }
We wish to study strings
propagating in the background \ffff\ with a curved transverse space.
This propagation is described by the
following \sm
\eqn\mof{  L_F=F(x) \del u \bd v +  (G_{ij} + B_{ij})(x)\ \del x^i \bd x^j
+ \a'{\cal R}\p (x)\ ,  }
where ${\cal R}$ is related to the worldsheet metric $\g$ and its
scalar curvature by  $ {\cal R} \equiv \fourth \sqrt \g R^{(2)}$.
We will refer to this model  as the `$F$-model'. We will also study the
following
generalization of the plane waves\eqn\mok{ L_K= \del u \bd v +  K(u,x) \ \del u
\bd u +  (G_{ij} + B_{ij})(x)\
\del x^i \bd x^j  + \a'{\cal R}\p (u,x)\ ,  }
which we will call the `$K$-model'.
When $K$ and $\p$ are independent of $u$,
these two models are simply related by leading order duality: the dual
of \mok\  with respect
to $u$ is \mof\ with $F=K\inv$.

The \FM has a large symmetry group. It is invariant under
the  three Poincare transformations
in the $u,v$ plane.  Moreover, it is invariant
 under
the infinite-dimensional symmetry $u\ra u + f(\tau+\s),
\   v\ra v + h(\tau-\s)$, i.e.  it has
two chiral currents. In general, the \KM has only one null Killing vector
$l= \del/\del v$, but it
is covariantly constant. The special case where $\p$ depends only on
$u$,
and
\eqn\plane{G_{ij} =\d_{ij}, \quad B_{ij} = 0, \quad K(u,x) = w_{ij}(u) x^ix^j}
actually has translation invariance in all transverse directions.
It can be put into the form $L_K= \del u \bd v +  {\tilde G}_{ij}(u)
\del
 x^i \bd x^j $.
The general \KM is also invariant under $v\ra  v + h(\tau-\s)$,
i.e. it has one chiral current.

The fact that the \KM has a covariantly constant null vector
can be used to give a simple geometrical argument that
   leading order solutions are  exact in the special case
when  $G_{ij} =\d_{ij}, \ B_{ij} =0$, and $\p$ depends only on $u$ \host.
This is because the curvature contains two powers of the
constant null vector $l$, and derivatives of $\p$ are also
proportional to $l$. One can thus show that all
higher order terms in the equations of motion vanish identically.
Only the leading order equations turn
out to be nontrivial. Can one extend this argument to the case \mok\
when the transverse space is nontrivial? Clearly, the curvature of
the transverse space  can now appear at all orders of $\a'$. Let us
suppose that the transverse space is  known to be an exact solution
in some scheme. Then the model \mok\ will be conformal with $K=0$. But
the curvature of the metric with $K \ne 0 $ is equal to the curvature
of the metric with $K=0$ plus a term of the form $(\na \na K)\ l\ l $.
Unfortunately, this can
result in nontrivial corrections to the equations of motion at
each order of $\a'$. These corrections will be linear in $K$, so one
learns that  the
exact equation for $K$ will also be linear.
But from this argument, one cannot
conclude that there is a scheme in which the leading order solution
for $K$ is also exact. To establish this, one needs to explicitly
study the conformal invariance conditions from the path integral
which we now proceed to do.

\subsec{ Generating functional  and conformal invariance conditions}
We shall study  the conditions of conformal invariance
of  the above models by directly looking at the  path integral representation
for the generating functional. To obtain  the complete set of
Weyl invariance equations  we need to introduce   sources
for the \sm fields and  find out when
the generating functional on a curved worldsheet
 does not depend on the conformal factor of the worldsheet
 metric. This is  equivalent to the condition of the vanishing of the trace
of the  stress-energy tensor operator.

To avoid duplication, we will start by considering a more general \sm which
includes both the \FM and the \KM\foot{Low energy solutions of this form
have been discussed in e.g. \garf.}
\eqn\mofk{ L_{FK}=F(x) \del u \bd v +  K(x,u) \ \del u \bd u  + (G_{ij} +
B_{ij}
)(x) \ \del x^i \bd x^j     + \a'{\cal R}\p (u,x)\ .  }
After proceeding as far as we can in this general theory, we will specialize
to the two cases of interest, the \KM ($F=1$) and the \FM
($K=0, \p = \p(x)$).\foot{By adding a multiple of $u$ to $v$,
 one can shift $K$ by a multiple of
$F$. So the theory \mofk\ with $K = bF$ ($b$ constant)
is also equivalent to  the $F$-model.
Note also that in the case when $K$ does not depend on $u$,
 setting $u=y_1+ y_2, \  v= y_1 + qy_2 $ and applying a duality transformation
in $y_2$ direction
 we find
($u,v\ra u',v'$)
$ F'= -{F/( K + qF) } \ , \ \  K' = -{1/( K + qF )} $\ . }
A slight generalization of the \FM\ which preserves its
conformal invariance is discussed in Appendix B.


Let us define the generating functional
\eqn\gen{ \exp (-W[U,V,X, \g])
=\int [\ du\ dv \ dx\ ]\ \exp \bl(
- {1\ov \pi \a'} \int  d^2z \ } $$ [\  L_{FK} (u,v, x,  \g)  +
V\del\bd u + U\del\bd v  +  X\del\bd x \  ] \br) \  , $$
where $U,V,X$ are  external sources on the worldsheet with metric
$\g_{pq}$
(which is taken in the conformal gauge).
Since $v$ only appears linearly in the action, one can do the integral over
it explicitly (by rotating $v\ra iv$),   obtaining a
  $\d$-function factor
\eqn\genn{ \exp (-W[U,V , X, \g])
=\int  [\ du\ dx\  ]\  \d\ [ \bd (F(x) \del u - \del U)] \  \exp  \bl(
- {1\ov \pi \a'} \int  d^2z\  } $$  [ \ K(x,u)\  \del u \bd u    + V\del\bd u
   + (G_{ij} + B_{ij})(x)\  \del x^i \bd x^j     + \a'{\cal R}\p (u, x) +
X\del\bd x   ]\br) \  .$$
The  $\d$-function now allows us to do the integral over $u$  to obtain
\eqn\gnnn{ \exp (-W[U,V, X, \g])
=\int  [\ dx\  ]\   \O [x,\g]  \  \exp \bl(
- {1\ov \pi \a'} \int  d^2z\ } $$  [ \  (G_{ij} + B_{ij})(x)\  \del x^i \bd x^j
  +  K(x,u_*)\  \del u_* \bd u_*   + V\del\bd u_*   + \a'{\cal R}\p (u_*,  x) +
X\del\bd x\   ]\br) \  ,  $$
where  $u_*$ is defined by  ($f$ is an arbitrary holomorphic function)
\eqn\ttt{ F(x) \del u_* = \del U + f(z)  \equiv  \del U' \ . }
The  determinant factor $\O$ is defined as follows. If the measure
for the $(u,v)$ fields is given by
\eqn\qqqq{ < \d u , \d v> = \int d^2z \sqrt \g  F_0 (x) \d  u  \d  v \ ,}
then
\eqn\ggg{ \O [x,\g] = (\det {Q})\inv \equiv  \int [du\ dv]\  \exp [-{1\ov \pi
 \a'} \int d^2 z F(x) \del u \bd v \  ] \ }
$$ \quad \quad \quad = \int [du\ dv]\  \exp [ - {1 \ov  \pi \a'}< u, Q \ v> ]\
,  $$
\eqn\qqqqb{  Q \equiv - {1\ov  \sqrt \g F_0}\del  (F \bd)\    . }
The crucial point is that $\ln \det Q$ can be computed explicitly
and  has a  $local$  form.
The general expression for $\det Q$
was found (in heat kernel regularisation)
in  \shwts\foot{A similar expression was also given  in \kalmor.
The reason why this determinant is given by a local expression
 can be understood in a simple way
by drawing an analogy with a complex scalar coupled to a $U(1)$ gauge field:
 $L= (\del \psi + i B \psi ) (\bd \psi ^* -i \B \psi ^*) $. If we set
$u' = \sqrt F\ u$ and $v' = \sqrt F \ v$, then the Lagrangian in \ggg\
takes the form  $L= (\del u' +
 B u') (\bd v' - \bar B v'), \ \ B=- \ha \del \ln F, \ \bar B= \ha \bd \ln F,
$
so that in the present case the gauge field potential is purely transversal,
${\cal F} = \del \bar B  - \bd B = \del\bd \ln F $.
The logarithm of the determinant is
proportional to that of a 2-dimensional Dirac fermion coupled to this gauge
 field, and
 the standard Schwinger-type
term in the effective action $\sim \int {\cal F} (\del\bd)\inv  {\cal F} $
is equal to
$\int \ln F \del\bd \ln F $. }
$$ \Delta I \equiv  - \ln\  \O
= {1\ov 8\pi  }  \int d^2 z  \sqrt{ \g } \Lambda^2 (\ln F- \ln F_0)
+  {1 \over 48 \pi } \int  R^{(2)}
\na^{-2} R^{(2)} $$
$$  -{1\ov 12\pi} \int d^2 z\  \big( \ \del{\ln } F  \ \bd  {\ln} F
+ \del{\ln } F_0  \ \bd  {\ln} F_0
+ 4\del{\ln } F  \ \bd  {\ln} F_0 \big) $$
\eqn\dee{
 -    {1\ov 24 \pi  }  \int d^2 z  \sqrt{ \g } R^{(2)}
 ( 2\ln F  + \ln F_0)  \     \ , }
where $\Lambda$ is an  UV   cutoff.

As usual, the  form of $\ln \det Q$ in \dee\ is not unambiguous being dependent
 on  a regularisation and
choice of measure, i.e.   is defined modulo local dimension 2 counterterms.
What is unambiguous is the  locality property of \dee.

The definition of the determinant, i.e. the
 choice of $F_0$  and regularisation  must be determined by
  the conditions on  the whole $(x,u,v)$ theory \mofk.
For example, if we use the  heat kernel regularisation and demand target space
covariance in the  $(x,u,v)$ space
we should set  $F_0=F$.\foot{For generic choice of measure/regularisation  one
may need to make a  non-covariant redefinition  of the dilaton field ($\p \ra
\p +\  a\  \ln \det G$)  in order to restore
the target space covariance of the model \mans. }
This is because the  covariant functional measure  for a \sm
with the  target space metric $G_{\m\n}$
is defined by
 $  < \d x , \d x'> = \int d^2z \sqrt \g  G_{\m\n}  (x) \d  x^\m  \d  x'^\n $.
Then  \shwts\ (we  do not indicate explicitly  the free-theory  $\g$-dependent
term in \dee)
\eqn\dddde{\Delta I = -{1\ov 2\pi} \int d^2 z \ \del{\ln } F  \ \bd  {\ln} F
 -    {1\ov 8 \pi  }  \int d^2 z  \sqrt{ \g } R^{(2)} \ln F \ , }
or, equivalently,
\eqn\dde{\Delta I = - {1\ov 2\pi} \int d^2 z\ \big( \ \del_i
  {\ln } F \  \del_j
\ln F\   \del x^i \bd x^j
 +    {\cal R} \  \ln F \ \big) . }
The two terms  in \dde\  lead to the local shifts of the metric $G_{ij}(x)$ and
dilaton
\eqn\shi{ G'_{ij} = G_{ij} - \a' X_{ij} \ , \ \ \  \ \ \p'= \p - \ha \ln F \ ,
\  \ \ \ X_{ij}=   \ha \del_{i}
  {\ln } F\   \del_{j}  {\ln} F \ . }
It should be  emphasized  that the  way  we compute the path integral \gen\
(by directly  integrating over $u,v$) does   not manifestly preserve  the
covariance in the   $(x,u,v)$ space. One usually  employs
 the normal coordinate expansion  in order to  maintain the
covariance of perturbation theory.
The
 use of \dee\ with $F_0=F$  in general is not  sufficient  to guarantee
the target space covariance of the full  theory.\foot{The covariance
 in the transverse $x$-space is  of course
 preserved  under proper choice of  regularisation/measure in the  $x$-theory.}
That means one may need to add extra  local $\del x \bd x $
{\it non-covariant}
 counterterms  in order to restore the target space covariance, i.e.
$X_{ij}$ in \shi\ may contain extra local terms constructed out of derivatives
of  $\p$ and $F$.\foot{One may question
why the addition of such  counterterms is  legitimate
given  that $\ln \det Q$ in \dee\ does not contain
 divergences of such kind.
 The point, however, is that
extra divergences may be present in the   general
$(x,u,v)$ theory which  should thus admit the  corresponding
freedom of local coupling redefinitions.}

Returning to functional integral \gnnn, we see that
since  $u_*$ is a non-local functional of $x$
the conditions of Weyl invariance of the resulting theory for $x^i$
are hard to determine  in a closed form.
We can, however,  proceed  further
in our two special cases of interest:

 (i) $F=1$\ \ \
and \ \ \ \ \  (ii) $ K=0, \ \p = \p (x) $.

\subsec { $K$-model  }
In the  \KM  \mok, $F =F_0=1$   so that
 the operator ${Q}$ \qqqqb\ is  trivial
and thus $u_* = U (z) $  is  $x$-independent (we absorb an arbitrary
 harmonic  zero mode  of  ${Q}$ in $U$).
The resulting  path integral is
\eqn\gnn{ \exp (-W[U,V, X, \g])
= Z_0 (\g) \int  [\ dx\  ]  \  \exp \bl(
- {1\ov \pi \a'} \int  d^2z\   [  \
  (G_{ij} + B_{ij})(x)\  \del x^i \bd x^j } $$ +   T(x,U)
    + \a'{\cal R}\p (U,  x) +
X\del\bd x  + V\del\bd U   ]\br) \  ,  $$
\eqn\tac{ T(x,U) \equiv   K(x,U)\  \del U \bd  U \ . }
Now it is easy to formulate the conditions of  the Weyl  invariance of this
theory:
(1) the transverse $x$- model  $(G_{ij}, B_{ij} , \p)$
must be Weyl invariant by itself (i.e., for $U=0$);
(2) since the  interaction potential $T$ is equivalent (in what concerns its
quantum field $x$ dependence)
 to a scalar  `tachyonic' term,
it should solve the `tachyonic' Weyl anomaly equation
which is linear (to all orders in perturbation theory
in $\a'$)  in $T$ \callnts. Since $T$ is proportional to $K$ we get
$$ -  \o T + \del^i \p \del_i T  + 2 \del^2_u  \p\  \del U \bd U =0 \ \ \   \ra
$$
\eqn\taac{  \
- \ha  \na^2 K   +  O(\a')  +  \del^i \p \del_i K+ 2 \del^2_u\p
=0 \ .  }
Here $\o$ is the scalar anomalous dimension operator
which in general contains $(G_{ij},B_{ij})$-dependent corrections to all
orders in $\a'$ and only  a few leading $\a'^n$-terms in it are known
explicitly (for a review  see \tspl).\foot{Note that  we have assumed that the
path integral
is computed in the  `minimal subtraction'
scheme where higher-order tadpoles  do not produce contributions to the Weyl
anomaly
so that the  operator $\o$  does not contain higher-derivative terms in the
flat space limit.}
The dilatonic terms appear due to the $x$ and $U$-dependence
of the dilaton.\foot{Computing the  variation over the conformal
factor of $\g$ one finds the `classical'  anomaly term  $\sim \del \bd \p$
which  (after use of classical equations of motion)
gives  $\na_i\na_j \p$ and $\na_u\na_u \p= \del_u^2 \p  + \ha G^{ij} \del_i K
\del_j \p$  in the relevant
terms in the operator of the trace of the stress tensor. }
Note  also that  in contrast to the usual tachyonic coupling, here $T$
has canonical dimension 2, so there is no `-2'   (tachyonic mass) term
in this equation. The  equation  \taac\ can also be interpreted as the
$uu$-component of the metric
$\bar \beta$-function of the original $D$ dimensional sigma model (see \tsnul).

Given an exact  string solution $(G_{ij},B_{ij}, \p )$,   in general,  we would
still be unable to determine the exact expression for $K$ because of the
unknown higher order terms in \taac.
There are, however, special cases when this is possible.
An obvious one is that of the flat transverse space with the  dilaton
 being linear in  the coordinate $x$
\eqn\ddd{G_{ij}=\d_{ij}\ ,\ \ \ \ B_{ij}=0 \ , \ \ \ \ \p = \p_0(u)  +
b_i(u)x^i
\ . }
Then the exact equation for $K$ \taac\  becomes
\eqn\tacc{
 - \ha  \del^i\del_i K  +  b^i \del_i K   +
2 \del_u^2  \p =0 \  }
and can be readily solved. For $b_i = 0 $  we obtain  the previously
discussed plane
wave type solutions \refs{\guv,\host,\desa,\tsnul}.
The special case where
\eqn\ffss{ K= 1 + {M\ov r^{D-4}}  \ , \ \ \ r^2 \equiv  x_ix^i \  , \ \ \
\p=\const}
is dual to the FS  background \hhs\ and describes a
string boosted to the speed of light.
For $b_i\ne 0$  one obtains a generalization of the plane wave type solutions
with a linear dilaton.

We  can obtain more interesting new exact solutions when the
CFT behind the `transverse' space solution $(G_{ij},B_{ij} ,\p)$ is
nontrivial but still known explicitly.
In fact, in that case the structure of the  `tachyonic' operator
$\o$ is determined by the  zero mode part of the
CFT Hamiltonian, or $L_0$-operator.
Fixing a particular scheme (e.g.,  the `CFT' one
where $L_0 $ has the standard Klein-Gordon form with the dilaton term)
 we are  then able, in principle,  to  establish  the form
of  the background fields $(G_{ij},B_{ij} ,\p)$
$and$ $K$. This produces a hybrid of a gauged WZW and plane wave solution.
Some examples in four and five dimensions will be discussed in Section 3.

\subsec{$F$-model }
Let us now turn to the second case when $K = 0$ and $\p = \p (x)$.
In that case  the substitution of $u_*$ in \ttt\  into the action in \gnnn\
gives
\eqn\gnt{ \exp (-W[U,V, X, \g])
=\int  [\ dx\  ]\     \exp \bl(
- {1\ov \pi \a'} \int  d^2z \  } $$
 [ \
 (G'_{ij} + B_{ij})(x)\  \del x^i \bd x^j      + T(x, U,V ) + \a'{\cal R}\p'
(x) +
X\del\bd x   ]\br) \  ,  $$
\eqn\tat{ T(x,U,V) \equiv -  F\inv (x) \del U'  \bd V  \ , }
where we have used \dde \  and $G'$ and $\p'$  were defined in \shi.
What we have obtained
 is a \sm for $x^i$ with the  `massless' couplings $(G',B,\p')$
and the `tachyonic' coupling $T$.
The dependence of $T$ on  the background sources $U,V$ only implies that  as in
the \KM\ \gnn\ and \tac,  $T$  has canonical dimension zero, not two.
The condition of Weyl invariance is thus that
$(G',B,\p')$ should represent a Weyl-invariant theory and $T\sim F\inv$
should satisfy  again eq.\taac\ (now with $(G',B,\p')$
as background fields)
\eqn\taccc{ -   \o' T +  \del^i \p' \del_i T
=0 \   \ra  \   - \ha  \na'^2 F\inv    +  O(\a')  +  \del^i \p' \del_i F\inv =0
\ . } As  mentioned  above,  this equation
can be written down explicitly to all orders in $\a'$
only  when   $(G',B,\p')$ corresponds to  a  known CFT.

To summarize, given a conformal  `transverse' theory  $(G', B', \p')$ and
$F$ satisfying  \taccc, we find  that  for the particular choice of couplings
in \shi\ the \FM with
\eqn\summm{  G_{ij} =G'_{ij}  +  \ha \a' \  \del_{i } {\ln } F \  \del_{j}
{\ln} F \ , \ \ \  \p= \p'   + \ha \ln F \ , \
\ \ B_{ij}= B'_{ij} \ ,   }
represents  an exact string solution (in a particular scheme
 the choice of which is implicit in the
definition of the path integral we were discussing).
Since the transverse theory  is,  in general,  defined modulo local coupling
redefinitions we may absorb the
$  \a'   \del_{i } {\ln } F  \del_{j}  {\ln} F$ term
in \summm\ into a redefinition of $G'_{ij}$ ($F$ is an extra scalar from the
point of view of the `transverse' theory).
We  may  also try to interpret this redefinition as a restriction
of a field redefinition in the full $(u,v,x)$ theory.

As we have already mentioned above,
the crucial point in our path integral argument is the locality
of the relation between $G_{ij}$ and $G_{ij}'$.  The precise form of this
relation would be fixed   would we  carry out  the argument using    some fixed
 explicit regularisation
of the whole $(x,u,v)$ theory. If  such a regularisation  does not manifestly
preserve
the target space covariance  we would need  to make local non-covariant
redefinitions
of the \sm couplings to restore the covariance in the final
expressions.\foot{Note that
from the point of correspondence with field redefinitions in the effective
action, the latter
ones need not necessarily be covariant in order to preserve the $S$-matrix.
The assumption of covariance is an extra  condition
that  restricts the class of effective actions and  field redefinitions  one
wishes to consider. From the quantum $2d$ \sm point of view, target space
covariance is an extra global symmetry in the space of $2d$ fields and
couplings
that needs special  effort (special choice of bare couplings)
 to be preserved in the full quantum $2d$  theory. }
The locality of \shi\ is sufficient in order to be able to  claim that there
exists a scheme where the \FM represents an exact   string  solution.
 We can use  the freedom of adding local non-covariant
counterterms to $X_{ij}$ in \shi\
to
 put  $ G_{ij} - G'_{ij}$ in a manifestly covariant background-independent form
which
is the  closest analog of \summm\foot{Similar redefinitions of the
metric appeared in the context of gauged WZW $\s$-models, relating a
`standard'  scheme to a scheme where the leading-order solution is exact
\refs{  \tspl,\tssfet} (see also  Section 6). }
\eqn\exaaa{  G_{\m\n} =G'_{\m\n} +  2 \a' \del_\m \p
 \  \del_\n \p \ , }
or simply  to remove $X_{ij}$  completely,
\eqn\exxx{ G_{\m\n} =G'_{\m\n} \ . }
This   can then  be considered as a  transformation
 to a `{\it leading-order}' scheme
where
  the transverse  metric is  given just   by the  $\a'$-independent $G'_{ij}$.

Let us now consider some examples starting again with  the simplest case:
\eqn\exa{  G'_{ij} =\d_{ij}\  , \ \ \ \ B_{ij} = 0, \ \
\ \ \p' = \p_0 + b_i x^i \ , \ \ \ \  \p_0, \ b_i =\const \ ,   }
or, in terms of the original fields  in the leading-order   scheme   \exxx
\eqn\exaav{  G_{ij} =\d_{ij}  \  , \
\ \  \p= \p_0  + b_i x^i   + \ha\  \ln \  F \ . }
In this  scheme, the exact  form of the equation for the function $F$ is
simply
\eqn\fff{ - \ha  \del^2 F\inv  +  b^i \del_i F\inv =0 \ , }
i.e. the  $F$-model
\eqn\fef{  L_{F}=F(x) \del u \bd v +  \del x^i \bd x_i    + \a'{\cal R}
 (\p_0 +b_ix^i +    \ha \ln F ) \  , }
with $F$ satisfying \fff\ is  Weyl invariant to all orders.\foot{The \FM \fef\
considered on a flat $2d$ background is thus UV finite on shell.
It may be
possible to prove this fact in a more direct way (without the
 need for an extra redefinition of the metric)  using the
 manifestly covariant normal coordinate expansion
in a way similar to how it  was done for the WZW model in \mukh.
One would then still have to show that there exists a dilaton such that
 the  condition of Weyl invariance (which is stronger than scale invariance
 \hutow)
is satisfied as well. }
In this scheme the leading-order duality is exact since the
leading-order  dual to \fef\ is the $K$-model
\eqn\feff{  L_{K}=\del u \bd v +   F\inv (x) \del x^i \bd x_i    + \a'{\cal R}
 (\p_0 +b_ix^i) \  ,   }
which represents an exact string solution if $F$ solves \fff\
(cf. \tacc\ with $\p =0$).
In particular, we conclude that there exists a scheme in which the FS  solution
\eqn\fsss{ F\inv ={ 1 + {M \ov r^{D-4}} }   \ , \ \ D>4 \ ;} $$  \ \
\ \ \ F\inv ={ 1 - \  {M \ \ln \   r } } \ , \ \ \  D=4\ , \ \ \ \ \
\ r^2 \equiv  x_ix^i \   , $$
is a classical string solution to all orders in $\a'$.

The conclusion   about the existence of  a scheme where the
$F$-model \fef\ represents an exact string solution  is consistent with the
result of \klts\  that
the particular $F$-model
\eqn\klt{  F\inv =  {  \sum_{i=1}^N \ep_i {\rm e}^{  \a_i\cdot x  } }\ , \  \ \
\ \
\p =   \p_0 +   \r \cdot x   +
\ha\  \ln \ F
\   ,   }
(where  the constants $\ep_i$ take values  $ 0$ or $ \pm 1$, $\ \a_i$
are simple roots of the   algebra of a maximally non-compact
Lie group $G$ of rank $N=D-2 $
 and $\r= \ha
\sum_{s=1}^m  \a_s$ is
 half  of the sum of all positive roots)
can be obtained  from  a $G/H$ gauged WZW
model. $H$ is  a  nilpotent subgroup of $G$
generated by $N-1$ simple roots (this condition on $H$
 is needed to get  models with one time direction). For example, the $D=4$
models
are obtained  for each  of the rank 2 maximally non-compact groups ($SL(3),
\ SO(2,2), $ etc.); for $\ep_i=1$,
$F\inv =  {\rm e}^{  \a_1\cdot x  } +  {\rm e}^{  \a_2\cdot x  } $ with the
classical string propagation being determined by the Toda equation.
 As argued in \klts, this background does  not receive  $\a'$
corrections in  the   `CFT'  scheme: since the gauged subgroup is nilpotent,
the  action or $L_0$ operator of the coset model is not modified by
$1/k$-corrections except for the standard  overall
rescaling $k\inv \ra (k + \ha c_G)\inv$ (see also Section 6).

\subsec{Remarks}
We have discussed  the \FM   from the point of view of
 a perturbative path integral approach.
One may try to give an alternative proof of its conformal invariance
using the   existence of the two chiral
 currents $\  J_u  = F\del u\   , \  \ \bar J_v =  F\bd v \ $
to construct directly the  conformal stress-energy
 tensor.  At first sight,
 it appears
  this idea should not work since  in contrast to the case of the
WZW  model
here  we do not have enough chiral currents.
The $x^i$-currents  do not look  chiral
since the $x^i$-equation of motion is not  free and has an interaction
potential proportional to $\del_i F\inv$ (for example,  this is the
 Toda equation in the case of the  models in \klt)
\eqn\qdqd{\del\bd x_i   =- \ \ha \ \del_i F\inv (x) \   J_u  (z) \bar J_v(\bar
z)\  .  }
 However, extra chiral currents may still exist.
This is illustrated by the example of the particular
$D=3$ \FM  with  $F= e^{-2bx}$  which is equivalent  (see Section 5) to the
$SL(2,R)$ WZW model and which thus must
have
extra  chiral currents in addition to $J_u$ and $\bar J_v$.
 In fact,  if  we define
\eqn\cuuu{
J_x = \del x + F v \del u = \del x  + v J_u \ , \ \ \ \bar J_x = \bd x + F u
\bd v  = \bd x + u\bar J_v\ , }
then the  classical equations  imply
\eqn\cuu{\del \bar J_v =0 \ , \ \  \bd J_u  =0 \ , \ \
\del \bar J_x =0 \ , \ \  \bd J_x =0 \ ,  }
where we have used the fact that
$F$ is a pure exponential.
Extra chiral  currents must  also exist for
the generic  $D=3$ \FM  since,
as we shall show  in Section 5, it is equivalent
to  a gauged $SL(2,R)\times R/R$  WZW model,
as well as for the the models \klt\  which  can  also be
obtained from particular gauged WZW models \klts.\foot{The  chiral
 currents in  $\s$-models  obtained by integrating out the 2d gauge field in a
gauged WZW model  should be non-local when expressed  directly
in terms  of the \sm fields.}

Another comment we would like to make is about possible
 supersymmetric generalisations. The model \mof\  has  an obvious $n=1$
 supersymmetric  version with  the fields $u,v,x^i$ replaced by $n=1$
superfields
(${\cal D}\equiv  {\del\ov \del \t}- \t {\del\ov
\del z}\ $)
\eqn\supa{  \int d^2z d^2 \theta  \   [  \  F(\hat x)\  {\cal D}\hat u \bar
{\cal D} \hat v +
(G_{ij} + B_{ij})(\hat x) \  {\cal D}\hat x^j \bar {\cal D} \hat x^i + \a' \hat
{\cal R} \p (\hat x)\   ]  \ . }
It would be interesting to formulate
conditions on the functions of the \FM under which
\mof\ admits   $n>1$  generalisations.
 Such extended supersymmetric versions exist for  the special $F$-models
which correspond to gauged WZW models.

As for generic $F$-models,  we can try to draw an analogy with
the case of the $n=1$ supersymmetric  gauged WZW  theories.
 In  supersymmetric WZW models   there is
no non-trivial shift of $k$ coming from the measure
since the Jacobian (${\hat y}$ and ${\hat y}'$  are  superfields of opposite
statistics)
\eqn\supam{ \int [d{\hat y}] [d{\hat y}'] \exp \{ - \int d^2 z d^2 \t\
  {\hat y} ( {\cal D} + [  A , \ ]){\hat y}' \ \}  \ \ ,   \ \ }
is trivial \tsnp: the fermionic and bosonic $A$-dependent contributions
cancel out.  This is the reason why   the corresponding
 \sm  couplings  receive no $\a'$-corrections  \refs{\jack,\bsfet,\tsnp}.
A similar conclusion is true for
the $n=1$ super-generalisation of the determinant factor $\O$ in \ggg\
\eqn\supan{ \O [\hat x ,\g] =   \int [d\hat u \ d\hat v] \ \exp \  [-\int d^2 z
d^2 \t \  F(\hat x) {\cal D}\hat  u  \bar {\cal D} \hat v\  ] \ .  }
Since
$F(\hat x) {\cal D}\hat  u
\bar {\cal D} \hat v = ({\cal D} + f ) \hat u' \ (\bar {\cal D} + \bar f )\hat
v' \ , \  \ f=  -\ha {\cal D} \ln F\ , \  \bar f=  -\ha \bar {\cal D} \ln F\ ,
$
it is natural to  expect that  this determinant factor contains only the
 dilaton contribution and  not the   derivative $f\bar f$-term
when defined in a supersymmetric way.
Then  the metric in \shi\  does    not have the $\a'$- correction.
This is certainly true for the particular $F$-models ($D=3$ and models in \klt)
 which are related
to  gauged WZW theories.

\newsec{New exact  solutions in five  and four dimensions}

We showed in the previous section
 that one could construct new exact solutions which
were a hybrid of the gauged WZW models and the plane waves by
using a gauged WZW model to describe the transverse space and adding
$du dv + K du^2$ to the metric.  The function $K $
must solve the  scalar `tachyonic' equation (with zero mass) in the transverse
space.
For a gauged WZW model, there exists a `CFT scheme'
 where the tachyonic equation is simple
(given by the zero mode part of the CFT Hamiltonian or $L_0$)
while $G,\ B$ and $\p$  may receive $\a'$ -corrections. By solving this
simple equation, and using the known exact form of $G,\ B$ and $\p$, one
obtains new exact solutions.

\subsec{Five dimensions}

The simplest example of this construction starts with the
$SU(2)$  WZW model. In this case the dilaton is constant, the metric
is  the standard round metric on $S^3$
\eqn\round{ds^2 = d\xi^2 + {\rm sin}^2 \xi\ d\Omega_2}
and $H_{ijk}=\epsilon_{ijk}$ is
the volume form. In
the CFT scheme, the metric and antisymmetric tensor
have only a constant overall rescaling and
the relevant equation  for $K$ is just
$\Delta K =0$, where $\Delta$ is the Laplacian on $S^3$.
Assuming $SO(3)$ symmetry, the  general solution is $K = a + m  \cot \xi$.
The constants $a$ and $m$ can be absorbed into a redefinition of $u$ and
$v$ so one obtains the following new exact solution:
\eqn\newfiv{ ds^2 = dudv + \cot \xi\  du^2 +  d\xi^2 + {\rm sin}^2 \xi\
d\Omega_2\  , }
with $H_{ijk}=\epsilon_{ijk}$ and $\p=$const as before. This solution has
singularities at the poles of the $S^3$. To
interpret these singularities, it is useful to consider the dual $F$-model.

Starting with the general solution for $K$ and dualizing with respect to $u$
 yields
\eqn\newff{ ds^2 = (a + m  \cot \xi)\inv  dudv +  d\xi^2 +
{\rm sin}^2 \xi\ d\Omega_2\ . }
Keeping the constants
$a$ and $m$ is necessary to obtain the general solution,
 since they correspond to the freedom
to dualize with respect to the symmetry which is a linear combination of
translations of $u$ and $v$.
In the \FM it is not possible to remove both $a$ and $m$
by redefining $u$ and $v$, but clearly
$a$ can be set to  be   either $\pm 1$ or $0$ by rescaling one of the
coordinates.
The dual
of \newfiv\ is the special case of \newff
\eqn\newfff{ ds^2 = \tan \xi\ dudv +  d\xi^2 + {\rm sin}^2 \xi\ d\Omega_2\ , }
Since $u$ in \newfiv\
is timelike on one hemisphere and spacelike on the other and so is  null at the
equator,  the
dual with respect to $u$ \newfff\ has
an additional singularity there.
 It
still has singularities at the poles, but in a neighborhood of these
singularities, the solution approaches the five dimensional
FS solution (see \fss)
\eqn\fivfs{ ds^2 ={r \over r+M} dudv + dr^2 + r^2 d\Omega_2\ . }
Thus even though the transverse space is curved, the singularities
introduced  by adding an $SO(3)$ symmetric $K$ in \newfiv\
 are just like the one in  the background  dual to
the FS. This is not surprising since locally the transverse space is,
of course, flat.

\subsec{Four dimensions}

To obtain a {\it four dimensional}
 solution one must start with a two
dimensional conformal $\s$-model. Essentially the only non-trivial  possibility
is
the   $SL(2,R)/U(1)$
gauged WZW model which describes the two dimensional euclidean black hole.
As discussed above, to construct this new solution we must
use all the zero-mode information provided by the
$SL(2,R)/U(1)$  coset: the exact metric and  dilaton   and the  form of the
tachyon equation. This will give us  the all-order form of all the
functions in the $D=4$  \sm  \mok. The solution  constructed in this way will
be
the $generic$ $D=4$ $K$-model.

Let us first review what is known about the exact background fields
of  $SL(2,R)/U(1)$ model in the CFT  scheme \dvv . The metric and dilaton
are given by
\eqn\exann{ ds^2 =  G_{ij} dx^idx^j= dx^2 + \ { {\tanh}^2 bx \ov 1 \ - {\ p \ }
{\tanh}^2 bx }   \ d{\t}^2   \ ,  \ \ \  }  \eqn\eeee{
 \p = \p_0  -  {\ln \ch }  bx   -
\fourth  \ln \bl( 1 \ - \ p \ {\tanh}^2 bx )\ ,  }
where the parameters $b$ and $p$ are related by
\eqn\exxxx{  \  p\equiv {2\ov k} \ , \ \ \ \ \ \
\a'b^2 = {1 \ov k-2}\ , }
$$D - 26  + 6\a'b^2 =  {3k\ov k-2} -1 -26  =0  \ , \ \ \ \ D=2\  . $$
Since $\t$ must be periodic, it is convenient to introduce
the shifted dilaton
\eqn\shtdil{\vp\equiv 2\p- \ha\ln  \det G\ . }
The physical coupling in this case is $\exp(\ha \vp)$ which is invariant under
the leading order duality transformation. For the above solution,
the shifted dilaton is simply
 \eqn\dii{ \vp = \vp_0 - \ln \sinh 2bx \ . }
Since in the CFT scheme the tachyonic equation has the
  standard uncorrected form,
 the  function $K(x)$ must satisfy (see eq.\taac; here $\p$ is
 $u$-independent)
\eqn\tach{ - \ha  \na^2  K  +   \del^i \p  \del_i K=
- {1\ov 2\sqrt G \e{-2\p} } \del_i (\sqrt G \e{-2\p} G^{ij} \del_j) K =  0 \ .
\  }
Observing that  the measure factor is $\sqrt G \exp (-2 \p) = c_0 \sinh 2r $
and $assuming$ that $K$ depends only on $x$ and not on $\t$
 we find that the solution of \tach\ is simply
\eqn\taa{ K= a  + \  m  \ \ln \tanh bx  \  . }
The constants $a, \ m $ can again
 be absorbed into  a redefinition of $u$ and $v$,
 so that the full exact $D=4$ metric is\foot{We can generalise this solution by
introducing a
$u$-dependent dilaton. Then $K$ will get an extra piece $\sim x^2$
and will grow at large $x$ as  in the plane wave case.}
\eqn\mix{ds^2 = dudv + \  \ln \tanh bx \ du^2  + dx^2 +    { {\tanh}^2 bx \ov 1
\ - {\ p \ } {\tanh}^2 bx }\ d{\theta}^2 \ ,  }
while the dilaton is unchanged.
This metric is asymptotically flat,  being a product of $D=2$ Minkowski
space with
a cylinder
at infinity.
Note that now the condition on $b$ is different than in the $D=2$ solution
since there are two extra dimensions
\eqn\cond{ 4 - 26  + 6\a'b^2 =  {3k\ov k-2} - 25 =0 \ . }

Remarkably, the solution for $K$ \taa\
 is the $same$ in the `leading-order' scheme
where the metric and dilaton do not receive $\a'$ corrections.
The point is that the tachyon operator remains the same  differential operator,
it is only its expression in terms of the new $G, \p$ that changes.
Thus, in the `leading-order' scheme we get the following exact $D=4$ solution
\eqn\mixl{ds^2 = dudv +  \ln \tanh bx \  du^2  + dx^2 +
   { {\tanh}^2 bx }\ d{\theta}^2 \ , }
$$ \p = \p_0  -  {\ln \ch }  bx \ , \ \ \  \vp = \vp_0 - \ln \sinh 2bx \ . $$
In addition to the    covariantly constant null vector $\del/ \del v$,
this solution  has
two
isometries  corresponding to shifts of $u$ and $\t$. Hence we can consider
two different types of duals. Dualizing with respect to $\t$ yields
\eqn\mixll{ds^2 = dudv +  \ln \tanh bx \  du^2  + dx^2 +
   { {\coth}^2 bx }\ d{\theta}^2\ ,  }
$$ \p = \p_0  -  {\ln\  \sinh \  }  bx \ , \ \ \  \ \  \vp = \vp_0 - \ln \sinh
2bx \ . $$
For the two dimensional euclidean black hole, this duality can be
viewed as a result of a coordinate
shift $bx \ra bx + i\pi/2$, under which the solution remains real.
This is no longer the case for the four dimensional solution \mixl\
since $G_{uu}$ is unchanged under this duality.

To obtain the dual with respect to $u$ we again need to start with the general
solution for $K$ \taa. We  get
\eqn\fmo{ ds^2  = F(x) dudv  + dx^2 +
 { {\rm tanh}^2 bx }\ d{\theta}^2 \ ,  \ \  \ \ B_{uv} = \ha  F(x) } $$
 \p = \p_0  -  {\ln \ch }  bx + \ha \ \ln\ F \ , \ \ \ \ F\inv = K = a+ m\ \ln
\th bx \ .   $$
In contrast to the euclidean $D=2$ black hole, the above  $D=4$
\KM and \FM metrics in \mixl\ and \fmo\  have curvature singularities at $x=0$.
The \FM\ \fmo\ may have an additional curvature singularity at nonzero $x$
 depending on
the parameters $a$ and $m$. Let us choose $a=1, \ m=-M \ (M> 0) $ so that these
additional singularities are absent. We thus find
\eqn\fmog{ ds^2  = (1 - M\ \ln \th bx)^{-1} dudv  + dx^2 +
 { {\rm tanh}^2 bx }\ d{\theta}^2 \ ,   } $$  B_{uv} = \ha (1- M\ \ln \th
bx
)^{-1} \ ,  \ \ \ \     \p = \p_0  -  {\ln \ch }  bx - \ha \ \ln\  (1 - M\ \ln
\th bx) \ . $$
The singularity at the origin of this solution is  exactly of the same type
that
appears in the    FS solution (see \fss):
 in $D=4$ we have $F\inv = 1 - M \ \ln\  r\  \ra -M\ \ln \ r $ near $r=0$,
 while here $ F\inv = 1- M\ \ln \th bx \ra  -M\ \ln \ bx $ near $x=0$.
As in the five dimensional example, the \FM\ \fmog\ is completely equivalent
to the fundamental string near $x=0$\foot{The
reason  why the two solutions agree is that near $x=0$ the dilaton is constant
and thus the equation for
$F:$  $\  - \ha \na^2 F\inv + \del^i \p \del_i F\inv =0 $
takes its FS form  $\na^2 F\inv =0$.}
\eqn\fmoo{x\ra 0\  : \ \ \  ds^2 \ra  (-M\ \ln\ bx)\inv dudv +
\      dx^2 +  (bx)^2 d\t^2  .}
This behavior near $x=0$ would be the same if we had started with the metric
\mix\ in the CFT scheme.

Moreover, the above \FM \fmog\   can be viewed as an improved version
of the FS solution  in four dimensions.  In $D=4$, the FS is given
by \ffff\ with $F\inv = 1 - M \ \ln\  r$. In addition to
the usual singularity at $r=0$ there is another singularity outside
the string at nonzero $r$. The solution we have just constructed
\fmog\ has the same singularity at  the origin (and hence can be viewed
as the field outside a fundamental string) but is regular elsewhere
and even asymptotically flat.
The original FS can be recovered by
taking the limit $b \ra 0$ which is consistent
since the central charge condition is now imposed
 only at the level of the full $D=4$ solution (and can be satisfied, e.g.,  by
adding 22
extra free degrees of freedom).


\newsec{Field redefinition ambiguity and structure
 of the effective action in $D=2$ and $D=3$}

The aim of this section is to discuss the general structure of the
tree-level string theory effective action (EA)
emphasizing a possibility to use the field redefinition freedom
to put higher order $\a'^n$-corrections in the simplest  form.
In particular, we shall show  that  the  EA
 can be chosen in such
a form (a `scheme') that all  $\a'$-corrections  $vanish$ once we specialise to
the case of a $D=2$ background.
In  such a  scheme the $D=2$  metric-dilaton EA is  thus known explicitly,
i.e. is given by the leading-order terms.
There also exists a scheme in which the $D=3$ limit of the EA
has all $\a'$-corrections depending only on the  derivatives of  the dilaton
but not on the curvature or antisymmetric tensor.

\subsec{`Scheme dependence' of the effective action}

Let us first recall a few basic facts about  the string  effective action
\refs{\scherk,\frts}.
Given a tree-level
string S-matrix  (in $D=26$) we can  try to reproduce  its massless sector
by a local covariant field-theory action $S(G,B,\p)$ for the metric,
antisymmetric tensor and dilaton. Subtracting the massless exchanges from the
string scattering amplitudes and expanding the massive ones in powers of $\a'$
gives an infinite series of terms in $S$ of all orders in $\a'$.

 The form of such action is  not  unique:
a class of actions
related by   field redefinitions  which are local,  covariant,
 background-independent, power series in $\a'$ (depending  on dilaton
 only through its derivatives not to mix different orders of string loop
 perturbation theory)  will correspond to the same string $S$-matrix.
Given some representative in a class of equivalent EA's  we refer to
other equivalent actions  as corresponding to different `schemes'.
The reason for this terminology is that the extremality  conditions  for  the
effective action are equivalent
to the conditions of conformal invariance of the
\sm representing string action in a background \refs{\lov,\call}  and the
related ambiguity in the \sm  Weyl anomaly coefficients or
 `$\b$-functions' can be interpreted
as being  a consequence of
 different choices of a  renormalisation scheme \tsred.
This implies that  some coefficients of the $\a'^n$-terms
in the EA will be unambiguous (being fixed by the string $S$-matrix) while many
others  will be `scheme-dependent'.

Though one possible way of determining  the EA is to start with  perturbative
massless string scattering amplitudes on a flat $D=26$ background,
 $S$  must actually be background-independent.\foot{After all,
we expect
the EA  to  be a result  solving for  the `massive
modes'
in  a hypothetical background-independent
string field theory action. } In particular, its  unambiguous coefficients are
 universal (e.g. they  do not dependent on the dimension $D$). This is implied
by the equivalence between the
effective equations of motion and the string \sm Weyl  invariance
conditions (which are  background-independent).
To  make this  equivalence  precise in any dimension $D$
we need only to  add       to the EA
one  $D$-dependent   (`central charge')
term $\sim \int d^D x \sqrt G \exp (-2\p) \  (D-26) $.\foot{There is
actually a subtlety
related to this term:  it is not clear that one can have a consistent
 $\a'$-perturbation theory if one is expanding near a vacuum with $D\not=26$.
For example, the linear dilaton background  will involve a parameter
of order $1/\a'$. This problem can be formally avoided by assuming that
 the central charge condition is imposed only at the very end and/or
 extra degrees of freedom are added to make total $D$ equal (or very close)
to 26. }

Given such a  background-independent EA
\eqn\act{ S = \int d^D x \sqrt G \  \e{-2\p}   \ \{ {2(D-26)\ov 3\a' }   -
  \  [ \   R \ + 4 (\del_\m \p )^2 - {1\ov 12} (H_{\m\n\l})^2\  ]
  + O(\a')   \}  \  , }
it  would be useful  to choose a scheme (i.e. the values of
  ambiguous coefficients) in which  $S$ has the
 simplest possible form.\foot{As usual in field theory,
one is trying to  fix  the freedom of local field redefinitions
in such a way that to have the simplest possible  action reproducing given
$S$-matrix. For example, one would prefer to  reproduce the graviton scattering
amplitudes
of the Einstein theory by the Einstein action but not  by  a  complicated
action $\int d^4 x \sqrt {G' }  R(G'), \ \ G_{\m\n}'
 = G_{\m\n}  +  \a' R_{\m\n}, $
which  contains all powers of the curvature.}
For example, the correspondence with the Weyl  anomaly coefficients of a string
\sm implies that there exists a scheme in which \act\ does not contain other
higher-order dilatonic terms.
This follows  also from the general argument \tsppl\ based on the path integral
representation for the EA \frts\ and was checked directly at the $\a'$-order
\refs{\metts,\jac}  by comparing with   string $S$-matrix. We now show that
in three dimensions one can do just the opposite, i.e., have only dilaton
terms as higher order corrections.

\subsec{Effective action in  $D\leq 3$}

It is possible   to
arrive at a more definitive conclusion  about a simplest possible scheme
by specialising to the low dimensional cases of
$D=2$ and $D=3$. More precisely, we would like to find  an  EA
 (defined for  general $D$)  such that  its $\a'^n$-terms take a simple form
in the limit $D\ra 2,3$.

 Given that   \act\ is background-independent (in particular,
its higher-order coefficients do not depend on $D$)
we are free to  take $(G,B)$
in \act\ to correspond to a generic $D=2$ or $D=3$ background.
Since the basic fields $(G,B)$ are second rank tensors, higher order
terms which involve `irreducible' contractions of tensors of rank greater
than two  cannot be altered
by field redefinitions. But
in $D\leq 3$   the Riemann tensor can be expressed in terms
 of the Ricci tensor, and $H_{\m\n\l} = \epsilon_{\m\n\l} H$. Thus
all possible covariant structures in the EA will have the `reducible' form
of products of scalars,  vectors, or at most,  second-rank tensors.

This is a necessary condition for a higher order term to be removed by
a field redefinition, but it is not sufficient. It has been shown \mettstwo\
 that some combinations of a priori ambiguous coefficients in
 the EA are actually redefinition-invariant
(unambiguous) and thus are uniquely  determined by the string $S$-matrix.
In fact, it is easy to show that one cannot  find  a scheme
in which there is no $\a'$-term in the $D=3$ EA. Suppose $H=0$ for
simplicity. Then
in the standard scheme, the $\a'$ correction to the bosonic string
EA is simply
\eqn\mettss { S_1(G,B,\p) =  a_0 \int d^D x \sqrt G \  \e{-2\p}
  (R_{\m\n\k\l})^2}
In three dimensions, $ (R_{\m\n\k\l})^2 = 4(R_{\m\n})^2 - R^2$.
Under a field redefinition, the action changes by a term proportional
to the leading order equations of motion. So if $S_1$ can be removed
by a field redefinition, it must vanish when the equations of motion
are satisfied (up to surface terms). Consider
first the Ricci term $\int  d^D x \sqrt G \  \e{-2\p} (R_{\m\n})^2$. Using
the low energy equations of motion  with $H=0$
and integrating by parts,
one can show that this is a total divergence in any dimension! But
the scalar curvature contribution $\int  d^3 x \sqrt G \  \e{-2\p} R^2$
turns out to be nonzero in general. Thus in $D=3$ the order $\a'$
term in the EA cannot be removed completely, although its form can
be altered. However in $D=2$, one can write
 $ (R_{\m\n\k\l})^2 = 2(R_{\m\n})^2$  so this term now vanishes
 when the leading order equations are satisfied. Using the results of
\metts\ one can show that this term
can indeed be removed by a field redefinition.

So far we have  discussed  just the first order $\a'$ correction.
What can one say more generally? Consider first the $D=2$ case
where $H_{\m\n\l}$ automatically vanishes.
Suppose we compute the scattering
amplitudes for the dilaton and graviton (in general $D$)
directly in the string frame where the dilaton and  graviton mix
in the propagator.
Since there are  no transverse degrees of freedom for the string in $D=2$,
there are no dynamical degrees of freedom in the $(G,\p)$ system,  and
the  limit $D \ra 2$ of the scattering amplitudes  is  trivial.
That means that  the on-shell limits of unambiguous terms in the EA
  must  vanish identically. Hence
 there exists   a choice of the EA (in generic $D$) such
that    higher order terms in
it  vanish in the $D\ra 2$ limit.

A similar statement is not true in $D=3$ since there is one transverse
degree of freedom for the string which could yield higher order
corrections to the scattering amplitudes and hence to the EA.
 However one can express
these corrections solely in terms of the dilaton. To see this, consider
the exact equations of motion for $G_{\m\n}$, $B_{\m\n}$, and $\p$
  in some scheme:
\eqn\eomgc{ R_{\mu\nu}  + {1\over 2}
         H^2 G_{\m\n}+ 2 \nabla_\mu \nabla_\nu
\p= \sum_{n=1}^\infty \a'^n T^n_{\m\n}\ , }
\eqn\eombc{ \nabla_\mu ( e^{-2\p}H )= \sum_{n=1}^\infty \a'^n V^n_\m\ ,  }
\eqn\eompc{ 4\nabla^2 \p - 4 (\nabla \p)^2 + R + {1\over 2} H^2
           - {2(D-26)\over 3 \a'} =\sum_{n=1}^\infty  \a'^n S^n \ , }
where $ T^n_{\m\n}$,  $V^n_\m$, and $S^n$ are the higher order correction terms
and we have used the fact  that $H_{\m\n\l} = \epsilon_{\m\n\l} H$
in $D=3$ (and assumed Minkowski signature).
Since we are interested in solving these equations perturbatively in $\a'$,
we can proceed as follows. Start with $n=1$ and use \eompc\ to replace
the $H^2$ terms in $ T^1_{\m\n},\  V^1_\m$, and $ S^1$ (and the left hand
side of \eomgc) by dilaton and curvature
terms. Then use \eombc\ to replace $(\nabla H)^2$ terms by the
 dilaton and curvature.
Finally, use \eomgc\ to replace all the curvature terms by derivatives of
 the dilaton.
This will, of course,  change the form of the correction terms for $n>1$ but
it will ensure that the
$n=1$ terms only involve the dilaton. One can now repeat this procedure
for each $n$. In this way, one can express all the correction terms
solely in terms of derivatives of the dilaton.
The action which reproduces this form of the exact equations
will then have only dilaton terms as higher order corrections.


\subsec{Discussion}

Let us discuss  some implications of the above  remarks. Since in $D=2$
 there is a scheme in which all $\a'$ corrections to the EA vanish,
all backgrounds which  solve  the leading-order equations
are in fact   exact solutions.
This conclusion is not  so surprising:  the $D=2$   `black hole'
 background \refs{\rabi,\witt} represents the generic solution
of the leading-order equations, and given that the corresponding CFT is known
($SL(2,R)/U(1) $ coset \witt)
one can find  explicitly \tsppl\  a local covariant background-independent
redefinition from the `CFT scheme'
 \dvv\   (where the background fields are $\a'$-
dependent) to the  `leading-order' scheme.
It also follows that in this scheme the $D=2$  \sm Weyl anomaly
coefficients  just have their leading-order form.\foot{One  may be
 tempted to draw a conclusion that there exists a scheme where
 the $\b$-functions of a generic
$D=2$ \sm also have just the leading-order form. This may not
necessarily  be the case since the field redefinitions  implied
 in our argument are more general (involving the dilaton)
than the  redefinitions corresponding to the freedom of choice of a
renormalisation scheme in the standard \sm $\b$-functions. It may be of
interest to understand this question further,
e.g.,
in connection with the RG flow in  some 2d $\s$-models  \onofri.  }

As for $D=3$,  in the scheme where $\a'$-corrections
are proportional to the derivatives of $\p$, the solutions
of the leading-order equations which have constant $\p$ remain exact
to all orders. It is easy to see that the only  leading-order solution
with constant $\p$ in $D=3$
is the constant curvature anti-de Sitter  space with  the parallelising
$H_{\m\n\l}$-torsion corresponding to  the $SL(2,R)$ WZW model or its possible
cosets over discrete subroups (in particular, the $D=3$  black hole of
\horwel).
For solutions with nonconstant $\p$,  the best that we can hope for is to find
a scheme in which a particular
leading-order solution does not receive $\a'$-corrections. This was shown
\tssfet\ to be the case (to $\a'^2$ order) for the
charged black string background  \hoho, i.e.
$SL(2,R)\times R/R $  coset model.
In the next section we
 will find  that there  exists a scheme where the leading order solution
for the $D=3$
$F$-model \mof,\fff\
is also conformal  to the next order in $\a'$.
This will  provide a perturbative check of  the general path integral argument
of Section 2.4.

Another  implication
concerns an exact form of the abelian duality transformations:
leading-order duality \busch\ is the symmetry of the leading-order terms in the
EA \refs{\banks,\tsmpl} and thus  is the  $exact$ symmetry in the simplest
scheme in $D=2$.
In fact,  we have checked
directly that while for a  general $D$ there   does not exist a   scheme
in which the  leading order duality  remains a symmetry  at $\a'$-order
without been modified by the derivative $O(\a')$-term \tsmpl,
such scheme does exist in $D=2$.\foot{ In  the case of the  $D=2$ black hole
solution  this was observed in \tsppl.}

As for $D\geq 4$,  here the
 massless sector of string $S$-matrix is non-trivial
so  no simple scheme  should  be expected to exist.

\newsec{ $F$-model in three dimensions}


The first non-trivial example of the \FM is in $D=3$. If $b=0$, the equation
for $F$ \fff\ just says that $F\inv$ is a linear function of $x$. Since the
transverse space here is one-dimensional, we can absorb the two free parameters
and write the solution as
\eqn\thrw{ F\inv = x   \ , \ \ \ \ \p = \p_0 - \ \ha \ \ln\ x  \ . }
This is the formal $D=3$ analog of the FS solution \fss.
The corresponding dual $K$-model
\eqn\sooo{ds^2 = dudv + x du^2  + dx^2 \ , \ \  \ \ \  \p=\p_0 \ , }
describes a flat spacetime.

The general solution of \fff\
 with $b\not=0$ is\foot{We shall assume that $b<0$ so that $x\ra +  \infty$
corresponds to the asymptotically flat region.}
\eqn\thr{ F\inv = a  + m  \e{ 2bx} \ , \ \ }
\eqn\bbh{ \ \ \p = \p_0 + bx - \ha \ln ( a + m \e{ 2bx} ) \ .   }
(The solution \thrw\ is recovered in the limit $b\ra 0$ provided one takes
$a$ and $m$ to infinity such that $a+m$ and $mb$ are kept fixed.)
This solution is closely related to the $SL(2,R)$ WZW model.
One way to see this is to note that in the limit $a\ra 0$, the \FM \mof\
 becomes equivalent to  the $SL(2,R)$ WZW model written
in the Gauss decomposition parametrisation
\eqn\maaa{  g =
 \left(\matrix{1&u\cr 0&1\cr }\right)
\left(\matrix{{\rm e}^{r}  &  0\cr    0  & {\rm e}^{-r}    \cr }\right)
\left(\matrix{ 1 &  0\cr  v  & 1 \cr }\right) , \   }
\eqn\sld{   L_{wzw}  =  {k} \big( \del r \bd r
  + {\rm e}^{-2r}  \del u \bd v  \big)    \ , \ \ \ \  r=bx + \ha \ln \ m \ , \
\ \a' b^2 = 1/k \ .  }
As we have noted earlier,
the actual value of $a$ is not physical since one can rescale $F$ by
simply rescaling one of the coordinates $u$ or $v$. The values
 which  yield  geometrically different solutions are $a=0, \pm 1 $.

Another connection between the \FM and the  $SL(2,R)$ WZW model is through
their duals.
The  \KM  dual to \thr\ is
\eqn\mokkk{ L_K= \del u \bd v + (a+m \e{ 2bx} ) \ \del u \bd u +
 \del x  \bd x + \a'
{\cal R}( \p_0  + bx) \ .  }
By the coordinate transformation $v\ra v-au$ and a rescaling of $u$ and $v$
 this becomes
\eqn\mokk{ L_K= \del u \bd v +    \e{ 2bx} \ \del u \bd u +  \del x  \bd x +
\a'{\cal R}( \p_0  + bx) \ ,   }
which is   obviously $u$-dual to the  $SL(2,R)$ WZW model \sld.
 In other words, the \FM  \thr\ is related to
 the  $SL(2,R)$ WZW model by dualizing with respect to one symmetry
and dualizing back with respect to another symmetry.
This implies that the  $D=3$ \FM \thr\  is an  $O(2,2)$ rotation
  of the  $SL(2,R)$ WZW model \sld.\foot{In any dimension $D$,
  the \FM with function $F$ is  related to the
\FM  with ${F'} = (F\inv + a)\inv, \ \ a=\const $  by the O(2,2) rotation (with
the dilatons
 being the same).
 Obviously, if $F$ is a solution of \fff\ the same is true for $F'$.}
This does
not prove that the \FM is equivalent to the  $SL(2,R)$ WZW model
since dual models are equivalent only if the symmetry direction is
compact; but if $u$ is periodically identified, the coordinate transformation
$v\ra v-au$ is not globally valid.

We shall  now show that the \FM with $b \ne 0$
 \thr\ can, in fact, be derived from an  $SL(2,R)\times
 R/R $ gauged WZW model. Let us first note  that the standard lorentzian
$D=2$ black hole \witt\
can be obtained by gauging the following global symmetry of the $SL(2,R)$ WZW
action in the Gauss decomposition parametrisation \sld:
$\ r'=r + \ep, \ u'= e^{\ep} u, \  v'= e^{\ep} v $.  The gauged action is
\eqn\wwss{   L_{gwzw}  =  {k} \big[(\del r  +  A)(\bd r + \A)
  + {\rm e}^{-2r}  (\del u  + Au) (\bd v  + \A v )  \big]\ .  }
Fixing the gauge $r=0$ and solving for $A,\A$ first  we finish with
\eqn\lrbh{L_{bh}=  k{\del u \bd v \ov 1+ uv} - \ha  {\cal R}\ \ln (1+ uv) \ .}
In contrast to the \FM   here one  cannot  easily integrate over $u,v$.
An equivalent expression is found by fixing the gauge as $uv=1, \ u= e^t$.
The resulting metric is then  given by
$ds^2 = k(1 + e^{2r})\inv (-dt^2 + dr^2)$.

 Introducing an extra field $y$ and gauging  independently
the  `left' and `right' subgroups of $SL(2,R)\times R $ (generated by the
positive and negative roots
as in \klts, i.e.,  corresponding
to the shifts of $u$ and $v$ in \maaa)
   we get  (cf. \wwss)
\eqn\gwz{ L_{gwzw}=   {k}\   \big[ \  \del r \bd r
  + {\rm e}^{-2r}  (\del u +  \l  A) (\bd v +   \n  \bar A ) \big]
  +  k (\del y +  \r A ) (\bd y +  \r \bar A )\     \ .     }
Here
the constants  $\l,  \n , \r $ correspond to a selection
of a particular subgroup
we are gauging (the action is invariant under: $ u'=u - \l \epsilon, \
v'= v-  \n \epsilon, \  y'= y -  \r \epsilon, \  A'=A + \del \epsilon, \ \bar
A'= \bar A + \bd \epsilon )$.
Fixing $y=0 $ as a gauge  and  solving for  $A,  \bar A $ we finish with the
$F$-model  (see also Appendix C)
\eqn\mode{    F\inv  =  a  +  \e{ 2r}
  \ , \ \ \ \ \phi = \p_0 +  r + \  \ha\   \ln\  F   \ ,
\ \ \  \ \ \  a  \equiv  {\l  \n \ov  \r^2}  \ .  }
This model is equivalent to \thr\  under the same identification as in \sld:
$r=bx + \ha \ln\ m , \ \a'b^2 = 1/k$.
Given the freedom of rescaling  $u,v$ and shifting  $r$
the only non-trivial values of  $a$ are  again  $0, +1, -1$.
$a=0$ (i.e. the limit $\r=\infty$   or    $\l =0$ or $ \n=0$)
 gives back the $SL(2,R)$ WZW model.

Gauging the  subgroup   of $SL(2,R)\times R $ which is a straightforward
extension of the  black-hole one in \wwss\  ($\ r'=r + \ep, \ u'= e^{\ep} u, \
v'= e^{\ep} v, \   y'= y -  \r \epsilon$) and fixing the gauge $u=v\inv=e^t$
 one can show that the corresponding   $SL(2,R)\times
 R/R $ gauged WZW model yields the following charged black string background
  \hoho\
$$  ds^2 = - \left(1-{M\over r'}\right) dt^2
           + \left(1-{Q^2\over M r'}\right)dy^2
           + \left(1-{M\over r'}\right)^{-1} \left(1-{Q^2\over Mr'}\right)^{-1}
                        {k \, dr'^2 \over 4 r'^2}, $$
\eqn\cbs{
B_{yt} = {Q\ov r'} \ ,  \ \ \  \ \phi = \p_0 - \ha \  \ln \  r'  \  , \ \  \
\p_0 = - \fourth \ \ln \ k \ ,  }
where $M$ and $Q$ represent the charge and mass per unit length. This
gauged WZW model only yields solutions with $Q<M$.  However, given
\cbs, one can clearly take the extremal limit  $Q=M$ to obtain\foot{The
 parametrisation of the charged black string background used in
\refs{\sfexac\tssfet}
$$ ds^2 = -{z-q-1\ov z } dt^2 + {z-q\ov z } dx^2
 + {dz^2 \ov 4(z-q-1)(z-q)} \ ,
 $$
is related to \cbs\  by $ z=\sqrt k r', \ M= q/\sqrt k  , \ Q^2= {{q(1+q)}/k},
\
$ so that  the extremal limit corresponds to $ q\ra \infty, \ k \ra \infty, \
M=Q=q/\sqrt k=$fixed.  Let us mention also that in  the Euler angle
parametrisation of $SL(2,R)$
$ g= {\rm e}^{{i\over 2 } \t_L \s_2 } {\rm e}^{\ha
{\tilde r}\s_1} {\rm e}^{{i\over 2 } \t_R \s_2  }, \ \t_L= \t + \tt, \ \t_R=
\tt-\t, $
the black string metric   is  \tssfet\
$$  ds^2 =
\fourth d{{\tilde r}}^2 +  (1 + q )  {C-1 \ov C+ 1+2q + 2b } d\t^2
  -q {C+1 \ov C + 1+ 2q } d\tt^2 \ ,  \ \  C= \cosh {\tilde r}\ .  $$
It  is  related to the above one in terms of $(z,x,t)$  by $  2z= C+1+ 2q\ , \
it=(1+q)^{1/2} \t\ , \ \ ix =-q^{1/2}\tt \ , $ i.e. $\t$, $\tt$ are to be
infinitely rescaled
 and   ${\tilde r}$ shifted in the extremal limit.}
\eqn\extre{ds^2 =   \left(1-{M\over r'}\right)(- dt^2 + dy^2)
                      +  {k \, dr'^2 \over 4 (r' - M)^2} \ . }
Letting $ bx $ denote the proper radial distance
\eqn\prodis{ bx = - {1\over 2} \  \ln\  (r'-M)\ ,   }
one finds  that the extremal black string
 is precisely the \FM  \mode\ or \thr\ where
the parameters are related by $a=1$,
 $m=M, \  \a'b^2= 1/k$. As we have just seen, one can  obtain this solution
directly as a gauged WZW model by  gauging a different subgroup of
$SL(2,R) \times R$.

The relation between the \FM and black string
clarifies   the causal structure of the  former. It was shown in \hoho\
that the extremal three dimensional black string \extre\ has a horizon
at $r'=M$ but no singularity. The correct extension across the horizon is
not to take $r'< M$ but to use a new radial coordinate $\eta ^2 = r'-M$.
The \FM in the form \thr\ just covers
the region outside the horizon and is incomplete.

In Section 2 we have found  that all $F$-models are exact solutions in some
scheme.
How this is consistent with the fact that $D=3$ \FM is equivalent to a gauged
WZW model?
First, there exists  a scheme in which the leading order solution for
the general charged black string \cbs\
remains a solution to the next order in $\a'$ \tssfet. In particular, this is
true in the extremal limit. The  above equivalence then
  implies  that the general
\FM in three dimensions
is  also exact to order $\a'$ in the same scheme.\foot{We have also
checked this directly  starting with the \FM and repeating the computation in
\tssfet.
We would like to point out a misprint in eq. (4.33) of \tssfet: it should
contain an extra term $-\na^2 S$. }

Moreover, by  taking the extremal limit of the  exact expressions
 for the  charged black string in the CFT scheme
\refs{\sfexac,\tssfet} one finds that there are no genuine $\a'$-corrections
in this case (all dependence on $1/k$  can be absorbed into rescalings of
the coordinates).
This is easy to see directly from \gwz.
The origin of the $1/k$ corrections to the \sm backgrounds
corresponding to gauged WZW models is in different renormalisation of the
coefficients ($k\ra k + \ha c_G$ and  $k\ra k + \ha c_H$) in front of the group
and subgroup parts of the action defined in the CFT scheme.  The CFT scheme
analog of \gwz\
thus has the coefficient  $k$ of the first $SL(2,R)$  two terms   replaced by
$k-2$
while the coefficient $k$ of the last  $R$-subgroup term remains
unrenormalised.
That means  that to find the exact background fields  the constant
$\r^2$  and thus $a$ in \mode\  are   to be replaced by $\r'^2= \r^2 {k\ov
k-2}$ and  $a'= a(1 - {2\ov k})$. But as was already  mentioned above, $a$ can
be rescaled  by a coordinate transformation.  We conclude  that like  the \FM
\klt\  obtained by the nilpotent gauging,
the $D=3$ \FM   does not receive non-trivial $\a'$-corrections not only in the
leading-order scheme but also in the  CFT scheme.

\newsec{Relation between  solutions in different  schemes}
Let us  now examine in more detail the relation between exact solutions in
different
schemes.  Since   $F$-models are in many respects similar to
gauged WZW models\foot{In particular, the integral over $u,v$ in the former is
similar to the integral over $A, \bar A$ in the latter.}   we shall start with
some general comments on
exact backgrounds corresponding to gauged WZW models.

\subsec{Solutions corresponding to gauged WZW models}
The  classical gauged WZW action  can be represented as
\eqn\gact{I_{gwzw} = kI_{wzw} (h\inv g{\tilde h})- kI_{wzw} (h\inv {\tilde h}),
\
\ \ \ \  A=h\del  h\inv\  , \ \  \bar A = {\tilde h}\bd {{\tilde h}}\inv  \ ,
}  i.e. as  a difference of  the two WZW actions for the total group $G$ and
the gauged subgroup $H$.
This representation  implies that the  gauged WZW  model is a  conformal
theory.
Fixing a  gauge on $g$ and changing the variables to $g' = h\inv g{\tilde h} ,
\
 h' = h\inv {\tilde h}$  we get a \sm on the group space $G\times H$ which is
conformal to all orders in a particular  `leading-order'  scheme.
That means that the 1-loop  group space solution remains  exact solution in
that scheme.
Replacing \gact\ with  the `quantum' action with renormalised levels
$k\ra  k + \ha c_G$ and $k\ra  k + \ha c_H$  does not change this conclusion.
This replacement   corresponds to  starting with the theory formulated in
the `CFT' scheme  in which, e.g.,  the  exact central  charge of the WZW model
is reproduced by the first non-trivial correction \refs{\tspl, \tssfet}
 and the metric $( k + \ha c_G ) G_{\m\n}$ is the one that appears in the CFT
Hamiltonian $L_0$  considered as a Klein-Gordon operator.

To obtain the corresponding  \sm in the `reduced'  $G/H$
configuration space (with coordinates being parameters of gauge-fixed $g$) one
needs to integrate out $A,\A$ (or, more precisely, the WZW fields $h$ and
${\tilde h}$).  This is a non-trivial step and the form of the result depends
on a choice of a scheme in which the original
 `extended' $(g,h,{\tilde h})$  WZW theory is formulated.

Suppose first   the latter is taken in the leading-order  scheme
with the action \gact.
Then the   result of integrating out $A,\A$ can be found by using a matrix
generalisation of the formulas \ggg,\dee,\dde.
If the $O(A^2)$ term in \gact\ is written as $\int d^2 z F_{ab} A^a\A^b$,
\ $  F_{ab}= \Tr (g\inv T_a g T_b - \d_{ab}), \ $ then  under a specific
assumption
about the measure
the  correction to the action is\foot{In  general, the derivative term in
$\Delta I$
will have the form $\Tr[\del f_1(F) \bd f_2 (F)]$ where $f_i$ are  functions of
the matrix $F$.  The choice of the measure should be consistent with  the
assumption that the resulting \sm should be
formulated in a  target space covariant way.
In particular, the resulting dilaton should be  the one
that can be also obtained by solving  directly the $covariant$ \sm conformal
invariance equations. }
 \eqn\ddem{\Delta I = -{1\ov 2\pi} \int d^2 z \ \del ( \ln \det F)  \ \bd  (\ln
\det F)
 -    {1\ov 8 \pi  }  \int d^2 z  \sqrt{ \g } R^{(2)} \ln \det F \ . }
The resulting \sm metric and dilaton are then given by (cf. \shi)
\eqn\resm{ G'_{\m\n} = G_{\m\n}  - 2 \a' \del_\m \p \del_\n \p \ , \ \ \
\p=\p_0  - \ha \ln \det F \ , }
where $G_{\m\n}$ is the metric obtained by solving for $A,\A$ at the classical
level
and $\p_0$ is the original constant dilaton.
Since the  $\a'$-term in the metric can be  eliminated by a field redefinition
we conclude that there exists a {\it leading-order scheme}  in which the
leading-order  gauged WZW
\sm
background $(G,B,\p)$ remains an  exact solution. The leading-order scheme
for the ungauged  WZW \sm   is thus   related to the leading-order scheme for
the gauged  WZW $\s$-model
 by an extra $2 \a' \del_\m \p \del_\n \p$ redefinition of the metric.
This   provides a general explanation for the observations in \refs{ \tspl,
\tssfet} about the existence of a leading-order scheme
 for  particular $D=2,3$ gauged WZW models.

If instead we start with the  $(g,h,{\tilde h})$  WZW theory in the CFT scheme,
i.e
 with the action
\eqn\gactt{I_{gwzw} = (k+ \ha c_G)\big[ I_{wzw} (h\inv g{\tilde h})- {k+ \ha
c_H\ov k+ \ha c_G}
I_{wzw} (h\inv {\tilde h})\big] \ , }
then the resulting  \sm couplings will explicitly depend on $1/k$
(and will agree with the  coset CFT operator approach results
\refs{\dvv,\bsfet,\tsnp}).
While in  the WZW model the transformation from the  CFT to the leading order
scheme is just a simple rescaling of couplings,  this  transformation  becomes
non-trivial    at the level of gauged WZW $\s$-model.   It is the  `reduction'
of the configuration space
resulting  from   integration over the gauge fields
$A,\A$  that is
 responsible for  a  complicated form of the transformation law between the
`CFT' and `leading-order' schemes in the gauged WZW $\s$-models  (in
particular,  this transformation  involves  dilaton terms of  all orders in
$1/k$, see  Section 6.2 below).

An exception is provided by the $\s$-models  \klt\ obtained by nilpotent
gauging:
here the second term in \gactt\ is absent by construction \klts. The background
fields do not receive non-trivial $1/k$ corrections even in the CFT scheme,
i.e.  the relation
between the leading-order and CFT schemes is equivalent to the one  for the
ungauged WZW model.
The  same is true for the $D=3$ \FM or the extremal limit of the $SL(2,R)\times
R/R$ coset.

\subsec{Transformations between different schemes }
 Let us now  discuss  how the above remarks are supported by the direct
perturbative analysis.
 There
exists  a simple (`standard')  scheme in which the order $\a'$ effective action
has the form \metts
\eqn\actt{  S = \int d^{D}x \sqrt {\mathstrut G} \ {\rm e}^{- 2
\phi} \big\{  {2(D-26)\ov 3\a' }   -
  \  [ \   R \  + 4 \na^2 \p - 4 (\na \p )^2 - {1\ov 12} (H_{\m\n\l})^2\  ]  }
 $$  - {1 \over 4} \a'\  [  \ R_{\m\n\l\k}^2 -\ha R^{\m\n\k\l}H_{\ \m\n}^\r
H_{\k\l\r} $$ $$
+ {1\ov 24} H_{\m\n\l}H^\n_{\ \r\a}H^{\r\s\l}H_\s^{\ \m\a}  - {1\ov 8}
(H_{\m\a\b}H_\n^{\ \a\b})^2\ ] +
O(\a'^3)   \big\} \  .  $$
The redefinition  leading from this `standard' scheme
to the the  leading-order   scheme
in which the  parallelizable (group)  space  with $\p=\const$
 is automatically a solution
of the conformal invariance equations (to order $\a'$) is  \metts
\eqn\rede{  G^{(lead)}_{\m\n} = G^{(stand)}_{\m\n} + \ha \a'  H^2_{\m\n}
+O(\a'^2)\ , } $$  \ \  \ \ B^{(lead)}_{\m\n} =
B^{(stand)}_{\m\n}  + O(\a'^2) \ ,
\ \ \ \p^{(lead)} = \p^{(stand)} + O(\a') \  ,    $$
where $ H^2_{\m\n} = H_{\m\a\b} {H_\n}^{\a\b}$.
As discussed above, this   scheme  should differ from the  leading-order scheme
in which the gauged WZW background fields do not receive $\a'$ corrections by
the dilatonic term in \resm.
In fact,  the  `leading-order'
scheme in which the  $[SL(2,R) \times R]/R $  gauged WZW  background
remains a solution at the $\a'$-order is  related to the `standard' scheme by
 \tssfet
\eqn\redd{  G^{(lead)}_{\m\n} = G^{(stand)}_{\m\n} + \ha \a'  H^2_{\m\n} - 2\a'
\del_\m\p \del_\n \p \  + O(\a'^2)
 \ ,  }
\eqn\reddi{
 B^{(lead)}_{\m\n} = B^{(stand)}_{\m\n}  + O(\a'^2) , \ \  \
      \p^{(lead)}=\p^{(stand)} + {1\ov 32}\a' (H_{\m\n\l})^2
     + {1\ov 8}\a' R  +
O(\a'^2) . }
The two $\a'$-terms in \redd\  thus  have  clear interpretation:
the first ($H^2_{\m\n}$) leads to the leading-order   scheme  for the ungauged
WZW model
while the second
$ (\del_\m\p \del_\n \p)  $  is related
       to  the  derivative term  \ddem\ in
the  determinant  factor   which
 results   from  the integration over the gauge fields in the gauged WZW
model.\foot{The transformation of the dilaton in \reddi\
does not seem to  have a simple interpretation  since we do not  explicitly
know how the dilaton is defined in the  `standard' scheme compared to the
leading-order scheme. }

At the same time, the  transformation
between the standard scheme and the CFT scheme (in which the background fields
receive corrections to all orders in $1/k$) was found  (for the $D=3$ $[SL(2,R)
\times R]/R $ background)   to be  \tssfet
\eqn\reddd{  G^{(cft)}_{\m\n}  = G^{(stand)}_{\m\n} + \ha \a'  H^2_{\m\n} -
2\a' (\del\p )^2\ G_{\m\n}  + \a' \na^2 \p G_{\m\n} + O(\a'^2)  \ ,  }
$$
 B^{(cft)}_{\m\n} = B^{(stand)}_{\m\n}  +   \a' \na^\l \p H_{\m\n\l} +
O(\a'^2)\ , \ \ \ $$
\eqn\reddii{
      \p^{(cft)}=\p^{(stand)} + {1\ov 12}\a' (H_{\m\n\l})^2  + {3\ov 8}\a' R  +
O(\a'^2)\ . }
The relation between the leading order  and CFT schemes obtained   in the case
of $D=2$
$SL(2,R)/R $ model  is \tspl
\eqn\cftf{   G^{(lead)}_{\m\n}  = G^{(cft)}_{\m\n}  - { 2\a' \del_\m \p \del_\n
\p  \ov 1  + \ha \a' R  }      + {2\a' (\del \p)^2  G_{\m\n}\ov 1 + \ha \a' R}
\ , }
$$  \p^{(lead)}=\p^{(cft)}  - \fourth  \ln ( 1  +
{\textstyle {1\ov 2}} \a' R  ) \ .   $$
Note that the presence of the dilatonic terms $\a' \del_\m \p \del_\n \p $
and $(\del \p)^2  G_{\m\n}$ in \cftf\ is consistent with \redd,\reddd.
However, since \cftf\ and \reddd\ were derived using specific properties of
$D=2$ and $D=3$ backgrounds, they need not coincide in detail.\foot{Since the
CFT scheme is defined only for specific gauged WZW backgrounds, there may not
exist a universal
relation (valid in  any $D$) between the CFT and the standard schemes. }

As follows from the  discussion in Section 5,
\redd\  is also the transformation to the leading-order scheme
for the $D=3$  $F$-model (see also Appendix C).  This transformation must be
universal:
it should  define the  `leading-order' scheme also  for generic $D> 3$
$F$-models.
This is certainly true for the models \klt\ and
 is consistent with what we have found in Section 2.
The background \exa\  corresponds to the solution in the `leading-order' scheme
\eqn\leadd
{G_{uv}^{(lead)} = B_{uv}^{(lead)}=\ha F \ , \  \
 \    G_{ij}^{(lead)}= \d_{ij} \ ,
  \ \ \ \p^{(lead)}
 = \p' + \ha \ln \ F \ , \  \ \p'\equiv  \p_0
+ b_i x^i \ .  }
We
thus conclude that   the result of the path integral argument of Section 2
is consistent with the
 perturbative analysis of the solutions corresponding to \actt\ (cf. \exaaa,
\redd).

\newsec{Discussion}

We have considered   two new classes of exact solutions to  bosonic
string theory. These take the form of the \FM \mof\ and the \KM \mok\ which
are related by a leading order duality transformation. One can view these
solutions as two different ways of extending a known spatial $D-2$ -
dimensional  euclidean CFT to
obtain a $D$-dimensional lorentzian one. The \KM provides an
interesting union of the standard gauged WZW and plane wave constructions.
We have discussed a four and five dimensional example in Section 3 but clearly
higher dimensional solutions can be constructed in an analogous
manner.
 Since these two classes of exact solutions are related by
leading order duality, it appears likely that  given any exact solution
to string theory
with a continuous symmetry, the solution obtained by a leading order
duality transformation is also exact in some scheme.

Perhaps the
most interesting solution in the class of $F$-models is the fundamental
string \fss.
This solution has a curvature singularity at $r=0$.  Furthermore,
the effective string coupling $ \exp \p = g_0 \sqrt F $
goes to zero at the singularity. The FS is the first example of an
exact solution with these properties.\foot{If one periodically identifies
the direction along the string, the effective coupling is $\exp(\ha \vp)$, $\
\vp = 2\p - \ha \ln G, $ which is invariant under  duality. Since the dual
of the FS is a plane wave which is known to be exact, this provides
a second (equivalent) example of a solution with these properties.}
It thus appears that this
singularity might survive not only   $\a'$-  but also
quantum string corrections. To establish this, two further results are needed.
One must  study the CFT corresponding to this
classical solution and
determine whether there is a singularity in a sense
appropriate to string theory. The diverging curvature of the metric `seen' by
point-like string states by itself is not
sufficient to ensure the existence of string singularities.
One must also study the string loop corrections in some detail.
The perturbative  corrections are powers of the string coupling $times$ some
(nonlocal) functionals of the metric and dilaton which may still  diverge
at $r=0$  producing a  large quantum  correction.
However, one should note that the string coupling vanishes faster in higher
dimensions while the curvature diverges like $1/r^2$  for all $D$.
Thus it is possible that perturbative
quantum effects are important only in low dimensions.
In addition, there may be   non-perturbative corrections, but
 they are likely to be small in the limit of small string coupling.

There are several arguments one might give to try to support the idea that
the singularity in the fundamental string solution
should be innocuous in string theory.
First, the behavior of classical test strings in this
 background has been studied
\cakh\ and it was shown that test strings parallel to the source string
and oriented in the same direction do not feel any force in the limit
of small velocities.\foot{This is modified if we make a periodic identification
to obtain strings of finite length $R$, and take the limit of small velocity
$v$ holding $Rv$ constant \ghrw.} This same conclusion holds for all
$F$-models. Second, quantum test strings have been studied
in a shock wave background  which has a singularity similar to the FS for
$u=0$,
 but is flat elsewhere \desas.
It was argued  that  string propagation remains well behaved.
Third, strings may be dual to five-branes,
and if one rescales the FS metric
by a power of the dilaton to obtain the geometry  seen
by a five brane, it does not have a curvature singularity \dgt\ (although
the dilaton still diverges). Finally, as we have said, the FS can
be viewed as the field outside a straight fundamental string.
Using the linearity of the equation for $F\inv$ one can consider the
multi-string solution and study string scattering. Preliminary calculations
show that this scattering is in agreement with the standard results of
string scattering in flat spacetime \refs{\cakh,\ghrw}.
This suggests that the FS solution is in some sense equivalent to the usual
strings in string theory which are certainly non-singular objects!

While it may be true that the singularity in the FS is not serious, the
above arguments
 are far from conclusive. In the first case,
generic classical string configurations certainly do feel a force, and
one must consider all states of a (quantum) test string before a singularity
is declared harmless. In the next argument, the fact that the spacetime
is flat away from $u=0$ means that a string will feel the singularity
for at most an instant. For the FS, the singularity is present for
all time. The third argument is relevant only if one wants to define a
singularity in terms of the behavior of objects other than test strings.
It is not clear whether this is a useful thing to do. Finally, the
scattering calculations have so far been compared only
  at large impact parameter where
the strong curvature regions do not play a significant role.
More importantly, we have
been viewing the FS as a particular nontrivial classical solution. The
question of whether it is singular is not directly related to the
scattering of two quantum strings. In particular, the fact that the
string coupling goes to zero in the FS solution has no analog in the
usual string scattering calculations in flat spacetime.

There is one unusual feature of the FS solution  which is evident even
at the level of the leading order string equations. While the FS is
certainly a solution to these equations for $r \ne 0$, it is not
a solution at the singularity
due to the presence of
 $\delta$- function source terms. In general relativity, one never asks if
the field equations hold at a singularity since in general this not a
well defined question. However if string theory is indeed a
`theory of everything', one should presumably not add external sources. One
might thus argue that the FS should not be viewed as an
allowed classical background. The difficulty with this argument
is that if one demands that the field equations hold everywhere, one is
in danger of simply defining away the problem of singularities. Physically,
one must study dynamical collapse situations to see whether the field
equations break down, or (in some sense) remain satisfied for all time.


The key property of the  $F$-models used in this  paper is their chiral
structure (the balance of the metric and antisymmetric tensor components) in
the $(u,v)$ sector.
That  is why the path integral over $(u,v)$ can be  computed exactly and one
can prove their all-order conformal invariance. The fact that the integral
over $(u,v)$ produces a $local$  effective theory for the  transverse
coordinates $x^i$
is quite remarkable. That means that for the $F$-model backgrounds one has
a formal  $D\ra D-2$   `dimensional reduction':
the correlators of operators  which  depend only on transverse
coordinates\foot{Unfortunately, most of these operators do not describe
physical states since a timelike momentum necessarily involves $u$ and $v$.}
are exactly given by the correlators in  the `transverse' euclidean CFT.
For example, for the $D=4$ $F$-models the `transverse' CFT is two-dimensional,
 i.e. it is either the  `flat space with linear dilaton'  (for the  FS or the
models \klt)
or the  $SL(2,R)/U(1)$ two dimensional  black hole  (for the solution
constructed in Section 3 \fmo). This is to be compared with, e.g., the
Schwarzschild  background where
integrating out any pair of coordinates produces a complicated non-local
two-dimensional effective theory.


We have seen in Section 5
 that the $D=3$ \FM can be obtained as a gauged WZW model.
It was previously shown that $F$-models of the form \klt\ can also
be obtained as gauged WZW models. An open question is whether
other  $F$- and $K$ -models, in particular, the FS one,
are also related to gauged WZW theories.
It is known that
some  of the
 plane wave solutions admit a  coset CFT  interpretation \napwietc.
 However, these are all of the  special form \plane\ and  thus are the simplest
type of \KM. It is not clear whether this correspondence extends to the
 more general $K$-models considered here.

Another interesting question  concerns  supersymmetric versions of the $F$-
and $K$-models,
and  related  superstring  and  heterotic string solutions.
In particular, we can construct a $D=4$ heterotic string solution by adding the
$(u,v)$-terms to the two dimensional  `monopole theory' constructed in \gps.
This is essentially a reinterpretation (in a Kaluza-Klein  or heterotic
string manner) of
  the  $(u,v) \times SU(2)$  $D=5$ bosonic solution
\newfiv\ as a $D=4$ heterotic one.


\newsec{Acknowledgements}
We wish to thank J. Gauntlett and K. Sfetsos for comments.
G.H. was supported in part by NSF Grant PHY-9008502 and by EPSRC
grant GR/J82041.
A.A.T. acknowledges the support of PPARC.


\appendix{A}{Leading order equations for the \FM}

We consider here the \FM with flat transverse space \ffff. Since
\eqn\add{G_{uv} = \ha F\  , \ \ \  G_{ij}= \d_{ij} \ , \ \ \ \
 \ B_{uv} = \ha F \  , \ \  B_{ij} =0 \ , \ \  \  F \equiv  \e{ 2 h}    \ ,   }
 the components of the  Christoffel symbols are
\eqn\rrr{ \Gamma^i_{uv} = - \ha F\del_i h \ , \ \ \
 \Gamma^v_{vi} = \del_i h \ , \ \ \
\Gamma^u_{ui} = \del_i h \ .  \ }
The only non-trivial components of the curvature  are
(others reduce to them  or vanish)
\eqn\aaa{ {R^u}_{iuj}= - \del_j \Gamma^u_{ui} - \Gamma^u_{ui} \Gamma^u_{uj} =
- \del_i \del_j h - \del_i h \del_j h\  , \ \ \
{R^u}_{uuv} = - \ha F \del_i h \del^i h\  .  }
The Ricci tensor is then
\eqn\rrrr{ R_{ij}  =- 2( \del_i \del_j h + \del_i h \del_j h ) \ ,
\ \   R_{uv} = - \ha F (\del^2 h  + 2\del_i h \del^i h)\ , \  }
\eqn\dddd{ R= -4 \del^2 h - 6 \del_i h \del^i h  \ . }
In addition, defining $ H^2_{\m\n} \equiv
  H_{\mu\lambda\sigma} {H_\nu}^{\lambda\sigma}$, one has
\eqn\bbbc{H_{iuv} = F \del_i h\ , \ \ \   H^2_{uv}= - 4F\del_i h \del^i h\  ,
\ \ \ H^2_{ij} = -8 \del_i h \del_j h\ , }
\eqn\moredd{  \na_i\na_j \phi = \del_i \del_j \phi\  ,  \ \ \
\na_u \na_v \phi = \ha F \del_i h \del^i \phi \ . }
The one-loop conformal invariance conditions can  be obtained
 by extremizing the action \act. Varying with respect to $G_{\mu\nu}$,
$B_{\mu\nu}$, and $\p$ yields
\eqn\eomg{ R_{\mu\nu}  - {1\over 4}
         H^2_{\mu\nu} + 2 \nabla_\mu \nabla_\nu \p= 0\ , }
\eqn\eomb{ \nabla^\mu ( e^{-2\p}H_{\mu\nu\rho}) = 0\ , }
\eqn\eomp{ 4\nabla^2 \p - 4 (\nabla \p)^2 + R - {1\over 12} H_{\m\n\l}
H^{\m\n\l}
           - {2(D-26)\over 3 \a'} =0 \ . }
Eq. \eomg\ yields
\eqn\ijj{(ij) :   \ \   -  \del_i \del_j h +  \del_i \del_j \phi =0 ,
\ \ \ \ \   (uv) : \ \  - \ha  \del^2 h + \del_i h \del^i \phi=0\  . \   }
Eq. \eomb\ does not  produce  any further
independent conditions. From the first
equation in \ijj\  one has $\p -h = \p_0 + b_i x^i$ where $b_i$ is a constant
vector. The second equation in \ijj\ can then be written
$ \del^2 F\inv = 2b^i \del_i F\inv$, and the dilaton equation
\eomp\ implies $b_ib^i = -(D-26)/6\a'$.

One might expect  the curvature to be simpler
using the connection with torsion
\eqn\sva{ \hat \Gamma^\l_{\m\n} =  \Gamma^\l_{\m\n} + \ha H^\l_{\m\n} \ . }
Then
\eqn\sqaa{ \hat\Gamma^i_{uv} = 0\ , \ \
\hat\Gamma^i_{vu}= - F\del_i h \ , \ \
\hat\Gamma^v_{iv} = 0\ , \ \
\hat\Gamma^v_{vi}= 2\del_i h \ , \ \
\hat\Gamma^u_{ui} = 0\ , \ \
\hat\Gamma^u_{iu}= 2\del_i h \ .  }
The curvature for $\hat  \Gamma^\l_{\m\n} $
is the following
\eqn\asa{ \hat {R^u}_{iuj} = -2 \del_i \del_j h \ , \ \
\hat {R^v}_{ivj} =0\ , \ \
\hat {R^i}_{ujv} = 0\ , \ \
\hat {R^i}_{vju} = - F  \del^i \del_j h \ , \ \ \hat {R^u}_{uuv}=0  \   .  \
}
It  vanishes when $h$ is  a linear function of $x$,
e.g. for the  $SL(2,R)$ WZW model  ($F=e^{-2bx} $) as it should
 since this
is the group space case. To obtain a vanishing
curvature for the general \FM
we would need to `add'
dilatonic  terms to  $ \del_i \del_j h$ and  $ \del^i \del_j h$
terms in \asa. It is not  clear  if this can be done in a systematic way
by modifying the connection.

\appendix{B}{A generalization of the \FM}

In this appendix we point out that there is a slight generalization
of the \FM which can also  be shown to be conformal using the arguments of
Section 2.
This is motivated by the following generalization of the \KM (see e.g.
\refs{\host,\tsnul}).
When the transverse space is flat  one can extend the \KM by introducing
an antisymmetric tensor background  of the form
$B_{iu} = B_i(x) $.
The \KM then becomes
\eqn\kkk{
 L_{K}=\del u \bd v +   K (x) \del u \bd u  +   B_i(x) (\del x^i \bd u
- \del u \bd x^i) +   \del x^i \bd x_i    + \a'{\cal R} (a + b_i x^i)
  \  .   }
In this case, $H_{\m\n\l} $ is
again proportional to the covariantly constant null vector and one
can show that all  terms in the conformal invariance equations for the \KM
which involve more than two powers of $H$ vanish identically. If we
define $ H_{ij } \equiv  2\del_{[i }  B_{j ]}$, these equations become (cf.
\tacc)
$$ \del^j (e^{-2 \p} H_{ij})   =0 \  , $$
  \eqn\nnn{  - \ha  \del^2 K    + b^i \del_i  K- \fourth H^{ij} H_{ij}
+  2 {\del^2_u  \p   }  +   O( {\a'}{}^s(\del^s H)^2 ) =0 \ .
}
The last equation equation still includes $\a'$ corrections but  these vanish
in the simple case of  a
`constant field strength'
 \eqn\siii{ b_i = 0 \ , \ \ \ B_i = - \ha H_{ij} x^j \ , \ \  H_{ij} =\const \
,
 \ \
 - \ha  \del^2 K   - \fourth H^{ij} H_{ij}
+  2 {\del^2_u \p  } =0 \ . }
When $K$ and $\p$ are both independent of $u$, one can construct a
generalized \FM which is dual to  this  solution.
Our conjecture in Section 1 implies that this
generalized \FM  should also be an exact solution.
We now show that this is indeed the case.

The Lagrangian  $u$-dual to the generalized \KM\ \kkk\  is
\eqn\ffk{
 L_{F}=F(x)  ( \del u  + B_i \del x^i) (\bd v  + B_i \bd x^i)
+    \del x^i \bd x_i  } $$   + \a'{\cal R}
 (\p_0 + b_i x^i  +   \  \ha \  \ln \ F ) \  , \ \ \  \ \ \  F= K\inv \  .   $$
This generalized \FM  is still chiral in the  $u,v$ directions.
Integrating over $v$  and $u$ we find  the following
expression for the generating functional (cf. \gnt)
\eqn\gnng{ \exp (-W[U,V, X, \g])
= Z_0 (\g) \int  [\ dx\  ]  \  \exp \bl(
- {1\ov \pi \a'} \int  d^2z\   [  \
   G'_{ij}  \del x^i \bd x^j} $$
+  B_i (x) (\del U' \bd x^i - \del x^i \bd V)
    -
 F\inv (x) \del U'  \bd V   +  \a'{\cal R}( \p_0   + b_i x^i )  +
X\del\bd x   ]\br) \  ,  $$
where $G'_{ij}$ and $T$ are the same as in      \shi\  and \tat.
 By power counting the conformal invariance conditions for the couplings in $W$
must be at most quadratic in $B_i$. We thus get back to  the conditions \nnn.
In the  leading-order scheme we thus obtain  the following
`constant field strength' solution
\eqn\conh{  b_i = 0 \ , \ \ \ B_i = - \ha H_{ij} x^j \ , \ \ \   H_{ij}=\const\
, \ \ \
   \del^2 F\inv  = -\ha  H^{ij} H_{ij}  \ . }
The solution that generalizes the FS one \fss\ is
thus
\eqn\genee{ F\inv ={ 1 - { H^{ij} H_{ij}\ov 4(D-2)} \  r^2  +
  {M \ov r^{D-4}} }
  \ , \ \ D>4 \ .}

\appendix{C}{Shifts of metric and dilaton in $D=3$ \FM}

To  clarify the
 meaning of the shifts of the metric and dilaton  implied by the  path integral
argument of Section 2, let us  consider first  a particular  example of a $D=3$
\FM  --  the $SL(2,R)$ WZW model \sld:
$F= e^{-2bx}, \ \p=\p_0, \ \a'b^2 = 1/k .$
 Using the    method of section 2 one can
demonstrate  explicitly its  all-order conformal invariance
and compute the {\it exact} value of  its central charge.\foot{A similar
computation of the central charge  for WZW models admitting the Gauss
decomposition parametrisation was discussed  in
\ger (see also \klts).}
In fact, integrating over $u$ and  $v$ we find according to \shi\
the following $x$-model:  $$  L= G_{xx}(x)  \del x \bd x  + \a' {\cal R} \p \ ,
 \ \ \ \ \p= \p_0  - \ha \ln \ F = \p_0  + bx \ , $$
\eqn\btg{ G_{xx} = 1 -  \ha \a' (\del_x  \ln \ F)^2 = 1 - 2\a' b^2 = {k-2 \ov
k} \  .  }
The integral over $(u,v)$ thus produces the
 effective renormalisation $k\ra k-2$ of the coefficient of the $ \del x \bd x$
term.
We have assumed that the measure factor $F_0$ in \dee\ is equal to $F=e^{-2bx}$
(as  implied by the Haar measure).
The conformal invariance condition \taccc\ is then satisfied
automatically and  the corresponding central charge  condition becomes
(cf. \exxxx)
\eqn\cev{ 0= D - 26   + 6\a' G^{xx} \del_x \p \del_x \p =
 D - 26  +  {6\a'b^2\ov 1 - 2\a'b^2 } = -26 +   {3k\ov k-2} \ .  }
This calculation  was  done in  the  leading-order    scheme
where the original  background fields do not receive  $\a'$ corrections.
Alternatively,  we  may start with  the corrected metric in \summm\ and
  after   having   integrated out
$u,v$ get  just $G_{xx}=1$.   In that scheme  the central charge equation is
thus
\eqn\ccd{ 0=D - 26  + 6\a' G^{xx} \del_x \p \del_x \p =
D - 26   +  {6\a'b^2} \ . }
To  obtain  the same exact expression for the central charge we need to
start with the WZW action in the CFT scheme (i.e. with $k\ra k-2$) so that
 $\a' b^2 $ is identified with $1/(k-2)$ (see also  \refs{\tspl,\tssfet}).

Let us  now   repeat this argument  for the  general $D=3$ $F$-model, i.e. the
gauged WZW model \gwz.
Changing the variables to $w,\tilde w$:  $A= \del w, \ \bar A = \bd \tilde w$
we have in the $y=0$ gauge
\eqn\gwzz{ L_{gwzw}=   {k}\   \big[ \ b^2  \del x \bd x
  + {\rm e}^{-2bx}  (\del u +  \l  \del w ) (\bd v + \n  \bd \tilde w)
  + \   \r^2  \del w \bd \tilde w \  \big]    \ .     }
This is the expression in the leading-order  scheme for the WZW theory. In the
CFT scheme
$k\ra k-2, \  \r^2 \ra \r^2 {k\ov k-2} $.  As was already noted above, this
transformation is trivial since it  can be `undone' by  rescaling of the
coordinates.
If we  first change  the coordinates
 $u\ra u'- \l w, \ v\ra v'-\n \tilde w$ then
\gwzz\  becomes the sum of decoupled actions  for  the $SL(2,R)$ WZW model and
the  free
$R\times R$ model for $(w,\tilde w)$. The theory is thus  obviously conformally
invariant.
Integrating over  $u'$,$v'$   and
$w,\tilde w$ we   get  the same resulting $x$-theory \btg, \cev\
  as in the ungauged WZW case.
 The integral over  $w,\tilde w$ gives only a constant  contribution { if}  we
assume
 that the measure factor $F_{0w}$ corresponding
to $(w,\tilde w)$ is trivial.

\def\tw{\tilde w}

Equivalent result should be found if
 we $first$ integrate over $A,\bar A$ or  $w,\tilde w$  (as we are supposed to
do
in order to  obtain  a \sm  corresponding to   a gauged WZW model).
We    can  compute the  resulting effective action by applying  \ggg,\dee\
to the integral over $w,\tilde w$  with  the action $\sim \int d^2 z F_w (x)
\del w \bd \tilde w$,
\eqn\fww{
F_w\equiv     1  +  a  {\rm e}^{-2bx}\ ,  \ \ \  \ a= \l\n/\r^2\ . }
  Taking
the corresponding  measure factor to be  equal to $F_{w}$
we find   (cf. \thr,\mode)
\eqn\mofs{  L_F= F \del u \bd v +
(1-  {1\ov 2}  \a'  \del \ln  F_w \del \ln F_w
)\del x \bd x   + \a'{\cal R}(\p_0 - {1 \ov 2}  \ln F_w )  \ ,  }
$$ F\inv =  a +   {\rm e}^{2bx} = {\rm e}^{2bx} F_w \ .
$$
 If we $now$ integrate over $u,v$   (assuming that  the measure factor for
$(u,v)$ is
$F_0= F$  as would be natural if we  would have started with
a \sm \mofs\ with the path integral measure defined by the corresponding
\sm metric)
we get the  $x$-theory  with (cf. \btg)
\eqn\ere{ G'_{xx} = 1-  \ha \a' \del \ln  F_w\del \ln  F_w
-  \ha \a' \del \ln  F \del \ln  F \ ,  }
\eqn\diu{  \p=\p_0 - \ha \ln F' - \ha \ln F  = \p_0  + bx \ .
 }
While the  dilaton \diu\
 is  the  same as in \btg\   the $\a'$-term in the metric  does not reduce
to the expected result
\eqn\expe{G'_{xx}
= 1-  \ha \a' \del( \ln  F_w + \ln F) \del( \ln  F_w + \ln F) =
1- 2\a'b^2 \ . }
The reason for this paradox lies in the fact that the redefinitions of $(u,v)$
should be consistent with covariance properties of the measure, i.e. they do
not, in general, preserve the covariance of the theory.
As in the path integral argument in  Section 2, extra local counterterms are to
be added to get a consistent result. In the present case we need to add a local
counterterm
leading to the `mixing' term
$- \a' \del \ln  F \del \ln  F_w$  in $G_{xx}$ in  \ere. Then the final result
 \expe\  is  the same as in the manifestly `conformal'  approach  when one
first
redefines $u,v$ to decouple them from $w,\tilde w$.

 More generally,  one can
consider the following  analog of the $(u,v,w,\tw)$ part of \gwzz:
\eqn\rrrr{ L= F_1 (\del u + \del w)( \bd v  + \bd \tw)
 + F_2 \del w \bd \tw \equiv F_{ab} \del u^a \bd v^b \ .  }
The two ways of computing this integral (by first integrating over $u,v$ and
then over $w,\tw$
or vice versa)  give equivalent results, i.e.  the relation\foot{$\Delta I_1$
and $\Delta I_2$  correspond to the $(u,v)$ and $(w,\tw)$  integrals and are
the same as in \dee.}
\eqn\relar{ \Delta I=  \Delta I_1(F_1) + \Delta I_2 (F_2) = \Delta I_1({F_1 F_2
\ov F_1 + F_2 }) + \Delta I_2 (F_1 +F_2) \ ,  }
is true only if the measure on the {\it full}   $(u,v,w,\tw)$   space
  is  consistently assumed to be the same in both cases.


\vfill\eject

\listrefs
\end